\documentclass[manuscript,screen]{acmart}

\setcopyright{acmlicensed}
\copyrightyear{2025}
\acmYear{2025}
\acmDOI{XXXXXXX.XXXXXXX}
\acmConference[Conference acronym 'XX]{Make sure to enter the correct
  conference title from your rights confirmation email}{June 03--05,
  2025}{Woodstock, NY}
\acmISBN{978-1-4503-XXXX-X/2018/06}

\acmSubmissionID{2092}

\usepackage{enumitem}
\usepackage[table]{xcolor}
\usepackage{longtable}
\usepackage{wrapfig}
\usepackage{subcaption}

\begin{document}

\title{Remotely Seeing Is Believing: How Trust in Cyber-Physical Systems Evolves Through Virtual Observation}

\author{Zhi Hua Jin}
\email{huazjin16@gmail.com}
\author{Kurt Xiao}
\email{kurtxiaoz@gmail.com}
\author{David Hyde}
\authornotemark[1]
\email{dabh@alumni.stanford.edu}
\affiliation{%
  \institution{Vanderbilt University}
  \city{Nashville}
  \state{Tennessee}
  \country{USA}
}

\renewcommand{\shortauthors}{Jin et al.}

\begin{abstract}
In this paper, we develop a virtual laboratory for measuring human trust.
Our laboratory, which is realized as a web application, enables researchers to show pre-recorded or live video feeds to groups of users in a synchronized fashion.
Users are able to provide real-time feedback on these videos via affect buttons and a freeform chat interface.
We evaluate our application via a quantitative user study ($N \approx 80$) involving videos of cyber-physical systems, such as autonomous vehicles, performing positively or negatively.
Using data collected from user responses in the application, as well as customized survey instruments assessing different facets of trust, we find that human trust in cyber-physical systems can be affected merely by remotely observing the behavior of such systems, without ever encountering them in person.
\end{abstract}



\begin{teaserfigure}
  \includegraphics[width=\textwidth]{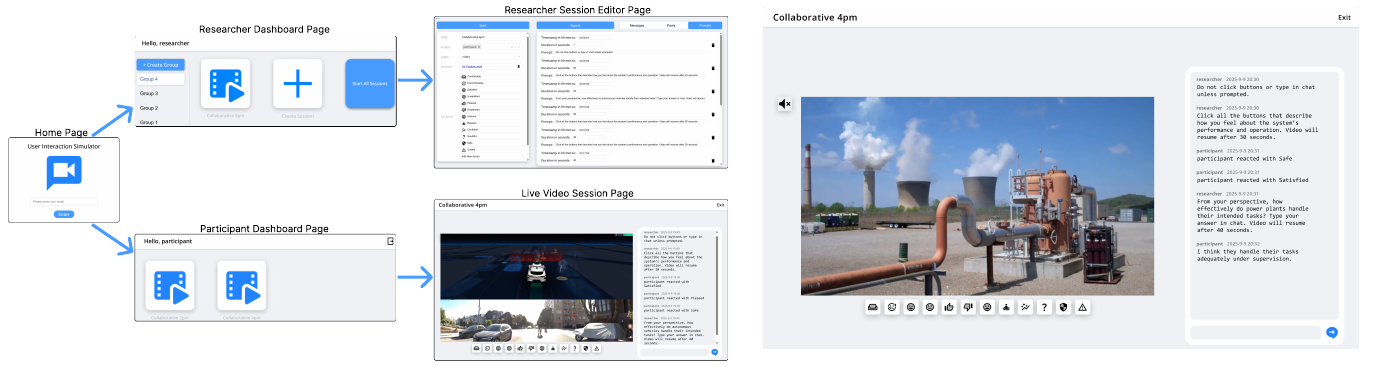}
  \caption{\textit{(Left)} A flowchart of using our virtual laboratory for trust measurement. After signing in to the web application, participants can join any study sessions to which they are added by researchers.  Researchers can create new study sessions, during which our interface allows specifying what video or content to display to participants, what affect buttons to show, and what prompts to ask participants (and when). Participants and researchers can then conduct and participate in a session, respectively. \textit{(Right)} A close-up of the study session view for a participant.  Here, the main portion of the screen plays a video of power plant operations.  Affect buttons below the video enable the participant to specify their reactions to content.  A freeform text chat interface allows users to answer qualitative questions from researchers, and to see other participants' reactions and responses.}
  \Description{On the left is a flowchart of different screenshots of our application, illustrating the user journey through our application for participants and researchers.  On the right is a screenshot of our ``virtual laboratory'' in operation.}
  \label{fig:teaser}
\end{teaserfigure}

\received{20 February 2007}
\received[revised]{12 March 2009}
\received[accepted]{5 June 2009}

\maketitle

\section{Introduction}
\label{sec:intro}
Cyber-physical systems (CPS) are network-connected devices and systems with physical presence or form.
CPS encompass a wide range of systems across scales of size and complexity; examples include autonomous cars, smart homes, pedometers, and smart grids.
Technologies in this category are fundamentally altering the human experience and the ways in which everyday citizens act and interact.
Given their importance, the study of CPS is an active area of research---including their modeling \cite{lee2015past}, design \cite{lozano2020literature,khaitan2014design}, and security \cite{kayan2022cybersecurity,alguliyev2018cyber}.
In the past few years, increasing attention has been given to human factors aspects of CPS, such as interface design and evaluation \cite{10.1145/3613904.3642958,10.1145/2851581.2892333,10.1145/3706598.3713889,10.1145/3706598.3714172,10.1145/3613904.3642356}.

As with any technology, the human factor of trust is essential---if not for the utility of a technology, then at least for its widespread adoption.
For example, trust (or lack thereof) in autonomous vehicles is well-known to influence their adoption \cite{lee2020exploring}.
Trust and acceptance are also known to be related in the case of industrial CPS, such as nuclear power plants \cite{keller2012affective,whitfield2009future,poortinga2003exploring,siegrist2000salient}.
Trust---which must be carefully defined and measured---has also been shown to be a dynamic quantity, and one that can be permanently altered when a CPS performs contrary to a user's expectation \cite{9304627}.

%Existing work on trust in the field of human-computer interaction (HCI) has predominantly focused on how trust initially forms and develops through \textit{direct} human-computer or human-system interactions.
%Prior studies have investigated how users' behaviors and trust dynamics shift in response to specific events or system behaviors during active user engagement tasks [].
%While interaction concerns the mutual determination between human and computer over time---where each influences the other's behavior []---our study explores an under-examined dimension of HCI: trust formation through passive observation.

In this paper, we study the development of user trust in CPS through a novel ``virtual laboratory'' for human trust.
Unlike conventional participatory studies where users directly manipulate, exchange information with, or control systems, our research investigates how trust develops when participants remotely observe CPS through curated video clips.
Our web interface enables users to provide feedback on videos of CPS they watch via affect buttons and a freeform textual chat interface.
The laboratory is configured so that a group of participants may all be shown the same videos simultaneously, and participants can see each other's feedback in real-time, mimicking a traditional focus group.

Our paper also presents customized survey instruments, based on established trust models and measurements typically used in direct interaction scenarios, to study trust formation in this passive observational context.
We use our virtual laboratory and these trust instruments to address the fundamental research question: how does merely remote observation of system performance influence participants' trust development over time?
Through a quantitative user study of approximately 80 participants, we find statistically significant evidence that virtual observation of cyber-physical systems performing well or poorly impacts human trust in the underlying cyber-physical systems.

% \begin{table}[H]
%   \caption{Research Questions Investigated in This Study}
%   \label{tab:research_questions}
%   \begin{tabular}{p{1.2cm}p{12cm}}
%     \toprule
%     \textbf{RQ} & \textbf{Research Question} \\
%     \midrule
%     RQ1 & How does passive observation of system performance influence participants' trust development over time? \\
%     RQ2 & What differences emerge in trust development patterns between participants who observe system performances collaboratively versus individually? \\
%     RQ3 & To what extent does peer presence during observation amplify or diminish participants' trust in the system? \\
%     \bottomrule
%   \end{tabular}
% \end{table}

\section{Related Work}
\label{sec:related-work}
Trust is an extensively researched concept in human-computer interaction (HCI).
From data visualization \cite{10.1145/3706598.3713824}, to social media \cite{10.1145/3613904.3642927,10.1145/3544548.3581019}, and of course, to AI and human-AI interactions and teaming \cite{10.1145/3706598.3713462,10.1145/3706598.3714286}, HCI researchers recognize the paramount importance---and in some cases, perils \cite{hardre2016and}---of trust in computer and digital systems.

However, as works like \citet{10.1145/3544548.3581019} and \citet{10.1145/3544548.3581197} point out, trust is not, on its own, a well-defined quantity, and one must rigorously define specific dimensions of and metrics for trust in order to conduct quantitative experiments and measure the effects of interventions.
In this section, after briefly situating the present work among the existing HCI literature, we focus on related work on defining and measuring trust in Section \ref{sec:rw-trust}, which motivates the interfaces and survey instruments we ultimately design and use for our quantitative user study.

\paragraph{Trust and Videos}
The present work performs interventions by showing real-time streaming or prerecorded online videos to users.
In their classic works, Bos and colleagues \cite{bos2001being,bos2002effects} assessed how interpersonal trust develops when users communicate with each other over videoconferencing (akin to our streaming modality) and find similar levels of trust as with in-person experiments, though they identified delayed trust development in the virtual setting.
These insights inspired our central research question, whether and how trust dynamics evolve when humans remotely observe cyber-physical systems rather than interact with other humans.
\citet{nguyen2007multiview} built upon \citet{bos2002effects} and found that eliminating spatial distortions in video feeds improved trust.
More recently, \citet{pan2016comparison} evaluated whether users trusted videos, robots, or avatars more when seeking expert advice.
\citet{horn2017exploring} considered how tutorial videos (e.g., in music) could build user trust.
In another line of inquiry, works like \citet{vaccari2020deepfakes} have studied the detrimental impact of deepfake videos on human trust.
Related to the present work, in \citet{avetisian2022anticipated}, the authors primed experimental subjects into low- and high-trust groups, and then showed them a video of autonomous vehicle performance and predicted emotional responses based on the priming.
In our study, we ask somewhat of the opposite question, using real-time measurements of emotion and pre-/post-survey instruments to assess how videos impact trust development.

\paragraph{Trust and Cyber-Physical Systems}
In the HCI literature, autonomous vehicles (AVs) are perhaps the most studied cyber-physical system (CPS) in recent years when it comes to human trust.
For instance, \citet{9304627} used a physical immersive AV simulator with a physical ``trust dial'' to measure how users' trust changed over time as the AV simulator behaved well or poorly.
In their study, where users physically interacted with (and at times, operated) an AV, the authors interestingly found that trust can be permanently damaged as soon as the AV performs poorly.
Along another dimension, \citet{10.1145/3706598.3713188} used machine learning to identify the most pertinent human factors that influence trust in AVs.
In a similar modality to the present study, \citet{10.1145/3411764.3445351} showed one video of an AV to participants across different groups, with each group seeing different semantic segmentation visualizations overlaid on the video.
Factors such as trust were measured with instruments including a German version of the Trust in Automation scale of \citet{jian2000foundations}.
Unlike \citet{10.1145/3411764.3445351}, our study shows videos of different CPS operating under differing conditions, and we leverage customized survey instruments, affect buttons, and freeform text answers to build our measurements of trust. 
On the modeling front, \citet{gil2019designing} developed a design theory for human-in-the-loop autonomous CPS, such as AVs, that aims to formalize many of the considerations found in human-AV and human-CPS trust papers.
Lastly, in another interesting work, \citet{rahman2019cognitive} presented a human-robot interaction framework where \textit{mutual} trust between the human and robot (CPS) is modeled and updated during the course of a task.

\subsection{Defining and Measuring Trust}
\label{sec:rw-trust}

%We discuss related trust literature across four key areas: (i) multi-dimensional trust constructs in HCI (ii) social influences on trust development (iii) trust calibrations (iv) instruments and methodologies for measuring trust, including validated questionnaires and emotions/keyword-based indicators.

%HCI frameworks have established trust as a multi-dimensional construct, identifying a variety of factors that influence trust formation in systems.
%Across the literature, researchers have proposed different models and terminologies, either tailored to specific subfields or in attempts to synthesize interdisciplinary definitions of trust.
%These variations have created challenges in establishing reliable, standardized instruments and consistently defined factors for trust research.
%To address this, [cite] compared, mapped, and analyzed the most frequently cited and best-validated trust survey instruments, ultimately proposing a structured model that includes three primary constructs, following []: dispositional trust, situational trust, and learned trust.

\subsubsection{Dimensions of Trust}
\label{sec:rw-dim}

To address and attempt to unify the diverse definitions of trust in the literature, \citet{10.1145/3677614} (see also \citet{doi:10.1177/0018720814547570}) compared, mapped, and analyzed the most frequently cited and best-validated trust survey instruments, ultimately proposing a structured model for trust that includes three primary dimensions:

\textbf{Dispositional trust} refers to an individual's general propensity to trust systems prior to encountering any specific situation or interaction.
%It reflects foundational beliefs and attitudes that shape a person's baseline approach to trust formation [all 3].
Although dispositional factors tend to persist over time, they can be influenced by long-term experiences, demographic variables, and personality traits \cite{10.1145/3677614}.
%These characteristics shape trust indirectly by influencing group membership perceptions, expectations of situation normality, and emotional responses [Hoff].
%Within the literature, disposition to trust has been further decomposed into distinct subdimensions. McKnight et al. (1998) propose two subconstructs: faith in humanity---the belief that others are generally competent, benevolent, and possess integrity---and trusting stance, a personal strategy of assuming trustworthiness in others regardless of evidence.
\citet{doi:10.1177/0018720814547570} extended the concept to human-automation contexts, framing dispositional trust as an individual's general tendency to trust technological systems.
More recently, \citet{10.1145/3677614} identified faith in technology as a key antecedent belief that influences expectations and emotional responses toward emerging technologies.
%Disposition to trust thus provides the broader framework for understanding how participants approach system observations, with faith in technology representing a key dimension particularly relevant for trust development through passive observation.
%As participants arrive with pre-existing tendencies that influence their initial interpretations of system behaviors, measuring dispositional trust prior to the observation experience allows us to distinguish between trust shifts attributable to inherent predispositions and those resulting from the media portrayals of system performance.

\textbf{Situational trust} reflects the extent to which trust is formed or adjusted based on characteristics of the immediate context.
%It is the attitude that an agent or system will help achieve an individual's goals in a situation characterized by uncertainty and vulnerability \cite{lee2004trust}.
Situational trust changes based on momentary environmental cues, user expectations, and conditions of the task at hand.
Users encounter these situations when they must rely on the incomplete knowledge of a system's intent and capabilities \cite{lee2004trust,wischnewski2024seal}.
Thus, in these contexts, situational trust emerges through the interaction between the trustor (the user) and the trustee (the system), with trust judgments shaped by perceived system performance, reliability, and contextual risk \cite{schultz2006trust,doi:10.1177/0018720814547570}.
%Razin et al describes three key antecedents that are widely recognized as influencing situational trust, encompassing one's shared mental model: Familiarity refers to the user's indirect knowledge of similar systems. Situation Normality refers to the alignment of a situation with expected norms of interaction. Emotional Response refers to the affective reactions a system and its behaviors evoke in the user. These do not represent trust themselves but act as critical precursors that shape how trust is initially formed and subsequently recalibrated as situations unfold.

\textbf{Learned Trust} refers to the trust that forms and adapts based on an individual's direct past experiences with a specific system or through ongoing system interactions.
Unlike situational trust, which is shaped by the immediate environment and contextual factors, learned trust develops from accumulated knowledge about a system's performance, behavior, and reliability over time \cite{10.1145/3530874}.
%Learned trust is divided into two forms: initial learned trust, which reflects the expectations users bring to an interaction based on prior exposure or reputation of the system, and dynamic learned trust, which evolves in real time as users observe and interpret the system's actual performance [Hoff and Bashir, 2015].
%Razin et al. (2024) further specify that General Trust and Intent to Use serve as key representative measures of learned trust.

%Disposition to trust thus provides the broader framework for understanding how participants approach system observations, with faith in technology representing a key dimension particularly relevant for trust development through passive observation.
The present work measures each of these dimensions of trust.
As participants in our study arrive with preexisting tendencies that influence their initial interpretations of system behaviors, measuring dispositional trust prior to the observation experience allows us to distinguish between trust shifts attributable to inherent predispositions and those resulting from the video portrayals of CPS performance.
Our study also leverages situational trust to examine how immediate environmental cues and observed system behaviors across varying scenarios shape participants' perceptions.
By evaluating key situational antecedents, we capture how users' initial trust perceptions are formed and how these perceptions recalibrate as new contextual information unfolds during passive observation.
Finally, we leverage learned trust to examine how participants' perceptions of the system evolve throughout the observation experience.
By assessing participants' initial confidence and tracking shifts in response to observed system behaviors, we capture both the influence of prior experiences and the development of new impressions during real-time engagement.

\subsubsection{Trust Calibration}
\label{sec:rw-tc}
In the user study discussed in this paper, participants observe CPS behaving under varying conditions over time.
Thus, participants undergo \textit{trust calibration}, which
\citet{MUIR1987527} defines as the process of aligning a user's trust with a system's actual capabilities, limitations, and behavior.
%---ensuring that trust is neither excessive (over-trust) nor insufficient (under-trust).
Rather than viewing trust as a fixed attitude or general belief about a system, Muir emphasizes that trust should be shaped over time through repeated observations of how a system behaves.
%Trust calibration, then, concerns not just how much someone trusts a system, but whether that trust is appropriate.
%Building on this foundation, \citet{lee2004trust} formalize calibration as a central component of appropriate trust, introducing a model that also includes resolution---how well users distinguish between varying levels of system capability---and specificity---how precisely trust is tied to particular functions or moments.
%Together, these dimensions help define when user trust in a system is appropriate: when individuals accurately reflect the system's actual behavior and capacity in a given context. 

\citet{10.1145/3544548.3581197} expanded this perspective by identifying distinguishing factors of trust calibration interventions, including several relevant to our work.
Our study performs endo calibrations---calibrations that take place during the interaction itself (rather than before or after, such as priming), allowing users to update their trust based on how a system performs in real-time.
Warranted calibration is relevant for CPS as it refers to strategies that aim to support trust judgments that accurately reflect the system's actual performance.
Warranted interventions are grounded in verifiable characteristics of the system, such as its accuracy or robustness, rather than relying on superficial cues like aesthetics or reputation.
Our study uses static calibration, which are interventions that are fixed and uniformly applied across all users (not adaptive based on user-specific behavior or real-time indicators of over- or under-trust).
%Instead, they provide the same information to all users regardless of prior experience or ongoing interaction patterns.
Finally, our study performs performance-oriented calibration, which involves focusing on what the system does---its reliability, accuracy, or task success---without necessarily revealing how those outcomes are produced.
This is particularly relevant for complex CPS such as power plants.
%These interventions present evidence of system effectiveness, enabling users to form trust judgments based on observable performance rather than internal processes. 
We refer to \citet{10.1145/3544548.3581197} for further details on the taxonomy of trust calibration.
%By distinguishing between when calibration occurs, whether it is grounded in actual system capabilities, how it adapts to user behavior, and what type of information it conveys, \citet{10.1145/3544548.3581197} offer a structured vocabulary for understanding how trust calibration operates in practice.
%This framing highlights trust calibration not only as a theoretical construct but as a practical tool for guiding user engagement---helping to prevent both overreliance and neglect of automated systems.

\subsubsection{Related Trust Instruments}
\label{sec:rw-ti}
%Prior work has emphasized the importance of using validated and supported instruments to assess trust in automated systems, particularly when trust is treated as a multidimensional construct.
Comprehensive meta-analyses and systematic reviews conducted by \citet{10.1145/3544548.3581197}, \citet{10.1145/3677614}, and \citet{10.1145/3530874} highlight the necessity of carefully aligning assessment instruments with their specific context of application.
Specifically, \citet{10.1145/3544548.3581197} advocated for the implementation of complementary methodological approaches that integrate both self-report measures and behavioral indicators to accurately capture fluctuations in user trust perceptions.
Furthermore, they emphasize the importance of trust assessment before, during, and after HCI.
Our study takes advice on both these fronts, conducting pre- and post-surveys as well as collecting behaviors and self-reported answers during our interventions.

Several established instruments capture different dimensions of user trust.
\citet{jian2000foundations} developed a scale to examine how system characteristics influence users' perceptions of trust.
They also noted that assessing individuals' general inclination to trust automated systems could serve as a valuable anchor for understanding how trust develops and changes over time.
Though well-established, \citet{jian2000foundations} does not factor trust into dimensions such as dimensional, situational, and learned trust.
Partially addressing this, \citet{schneider2017influence} created an instrument for measuring dispositional trust that captures stable cognitive and affective propensities toward system trust.
We use their instrument to assess dispositional trust in our study.

\citet{mcknight2002developing} presented a structured trust framework and accompanying questionnaire that distinguishes between disposition to trust, institution-based trust, trusting beliefs, and trusting intentions.
%This multidimensional model helps explain how trust develops both before and during system interaction.
We found their trust dimensions to be equivalent to the dispositional, situational, and learned factorization of trust discussed above:
%Their distinction between enduring personal tendencies and trust evaluations shaped by context aligns with \citet{10.1145/3677614} taxonomy, which categorizes trust as dispositional, situational, or learned.
disposition to trust corresponds to dispositional trust, institution-based trust is situational, and trusting beliefs and trusting intentions map onto learned trust.
%McKnight's disposition to trust corresponds to dispositional trust in Razin's framework, while institution-based trust reflects the situational factors that influence trust in a given environment.
%In parallel, trusting beliefs and trusting intentions map onto learned trust, as they describe how users form judgments and make decisions based on what they observe or expect from the system.
Although their instrument does capture different dimensions of trust, it is specifically designed for e-commerce, which made it difficult to directly adopt for our study on CPS.

For our purposes, \citet{lee2004trust} combines the advantages of \citet{jian2000foundations} and \citet{mcknight2002developing}.
The authors present a trust model focused on trust in automation (which is sufficiently similar to CPS) that also divides trust into various measurable dimensions that can be measured.
Their trust framework considers an assortment of descriptors that contribute to categorical characterizations of the performance (how reliably and competently the system achieves its goals), process (how understandable and appropriate its underlying operations appear), and purpose (the extent to which the system's intent and use align with the user's goals and values) of an automated system. 
We adapted this framework to construct our questionnaire and assess participants' learned trust.
%We based our assessments of learned trust on an adaptation of their framework.
Importantly, these three dimensions also underscore the formation of situational trust, providing the informational basis through which users recalibrate their trust in response to contextual cues about system performance, process, and purpose. We further derived our real-time qualitative trust questions from this framework, where each question corresponds to one of the three bases, enabling us to capture trust dynamically during CPS observation. 
%Something to end - not sure if the last sentence is sufficient, but I was thinking of tying it back to CPS and our videos in a better way? but i think this connects lee and see well to situational and learned now.

While many validated instruments---such as those developed by \citet{jian2000foundations}, \citet{mcknight2002developing}, and \citet{schneider2017influence}---explicitly reference trust in their item wording, our study intentionally avoids using the word altogether.
This decision reflects a design choice aimed at reducing priming effects and encouraging participants to respond based on their perceptions and judgments of the system's behavior, rather than aligning their answers with preconceived notions of what trust entails.
This allows us to explore how these underlying constructs develop through observation, without shaping participant interpretation through the language of trust itself.

% Jian doesn't divide trust into dispositional/etc.
% McKnight is specifically on e-commerce, so while it's factored, need to adapt

% Jessup is what we use for dispositional questions in our pre-survey -> find/replace technology with CPS

\section{Virtual Laboratory Design and Study Capabilities}
\label{sec:design}

Our study platform is a web application that connects participants and researchers in real-time sessions.
A researcher may create various sessions using our application, and subjects may or may not be invited to different sessions.

Each session involves watching a pre-recorded video or a live video stream.
During a session, subjects are able to interact via a freeform text chat interface\footnote{While not directly related to trust, we highlight the recent work of \citet{10.1145/3706598.3713682}, which evaluates the utility of real-time synchronized floating comments on videos.  This is similar to our system allowing participants to see each other's text chat messages and video reactions in real-time.}.
At specific times (chosen by the researcher when setting up the session), subjects are prompted to click affect buttons or to answer qualitative question prompts in the chat.
The inspiration for our platform was to build a system that enables workers or study participants to complete tasks that require synchronized effort and that may require collaboration (see, e.g., \citet{patent-app}); for example, one may wish for study participants to collaboratively analyze behaviors observed during a live video feed of a complex CPS.
These capabilities are in contrast to platforms like Amazon Mechanical Turk \cite{mturk}, which do not facilitate real-time or collaborative tasks or studies.

Though not utilized during the study in Section~\ref{sec:user-study}, we also built support in our platform for recording where subjects click within a video during the course of a study.
For instance, one could ask participants to watch a sports game, and to click when and where they observe fouls or rule violations.

Our virtual laboratory automatically records, organizes, and saves all data collected from users during a session into CSV files for later analysis.
All session users must authenticate to the web application with their email address, so the system can consistently associate logs and records with the correct user.

The laboratory's web frontend is built using React, HTML, and CSS.
Server-side infrastructure is built using Node.js, a JavaScript runtime environment; thus, both the frontend and backend are primarily in JavaScript.
Backend code is written using the MVC architectural pattern for ease of maintainability and scalability.
A standard MySQL database stores session and user information.
Finally, the Socket.IO JavaScript library is heavily used to enable bidirectional communication between clients and servers via WebSockets.
All infrastructure is hosted in the cloud (AWS) and can run using only small or medium instance types.

Figure~\ref{fig:teaser} illustrates the typical user flow through the web application (left), for subjects or researchers.
The right-hand side of Figure~\ref{fig:teaser} offers an example of the session view for a participant, including the video content, the affect buttons (see also Figure~\ref{fig:affect-buttons}), and the freeform text chat interface.

\section{Trust Instrument Construction and Trust Scores}
\label{sec:ti}

As motivated in Section~\ref{sec:rw-trust}, we design trust instruments---including pre- and post-surveys, as well as endo calibrations---that aim to measure dispositional, situational, and learned trust.
To measure dispositional trust, we present study participants with a pre-survey that is the dispositional trust instrument of \citet{schneider2017influence} (referred to in \citet{jessup2019measurement} as Propensity to Trust Technology) with the word ``technology'' replaced with ``cyber-physical system'' throughout.
(We note this is acceptable practice and similar to what \citet{jessup2019measurement}, for instance, did in their study, replacing ``the system'' with ``the runner'' in the instrument of \citet{jian2000foundations}.)

\begin{wrapfigure}{l}{0.4\linewidth} % "l" = left side, width = 40% of text width
    \centering
    \includegraphics[width=\linewidth]{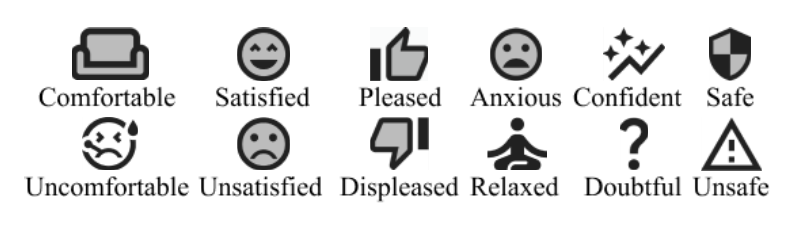}
    \caption{Affect buttons used in our study. Here, the buttons are arranged in bipolar pairs.}
    \Description{Two rows of the affect button pairs used in our study.}
    \label{fig:affect-buttons}
\end{wrapfigure}

Aside from the dispositional questions in the pre-survey, our pre- and post-surveys are identical, and we leverage them to assess learned trust relative to baseline values (i.e., we assess the difference in values between the pre- and post-surveys).
The surveys consist of 18 items based on keywords identified in \citet{lee2004trust} and well-corroborated by established trust research \cite{jian2000foundations,mcknight2011trust,korber2018theoretical,schaefer2016measuring,malle2021multidimensional,madsen2000measuring,chancey2017trust,wojton2020initial,gefen2003trust}.
Each item is measured on a standard five-point Likert scale.
For ease of exposition, we present each item, along with supporting quotes from the literature, in Appendix~\ref{sec:tos-survey}.

Regarding situational trust, we perform endo calibrations, prompting study participants to provide freeform text and affect button responses at specific times during a study.
Freeform text responses are collected via the chat interface in our application (see Section~\ref{sec:design}), whereas our web interface presents a pop-up for users to directly click on affect buttons.
For qualitative questions, we adapt our prompts from \citet{lee2004trust}.
For the affect buttons, the emotions we allow users to select are visualized in Figure~\ref{fig:affect-buttons}.
We adopt these emotions from the Geneva Affect Label Coder of \citet{scherer2005emotions} (see their Table~4).
Given the focus of the present study on CPS, we prune several categories that are unlikely be the dominant emotion a study participant would express towards a CPS during the study (e.g., longing, lust).
We also seek to ensure that each emotion appeared as a bipolar pair; for affect categories lacking a clear polar opposite in \citet{scherer2005emotions}, we consulted several highly-cited works \cite{bartneck2023godspeed,bradley1994measuring,osgood1957measurement,laugwitz2008construction} to add appropriate opposites.
Given the importance of safety to CPS, we also add safe/unsafe as an affect pair to our instrument.

In analyzing our results (Section~\ref{sec:results}), we seek to understand how overall ``trust`` evolves for study participants.
To aid in this analysis, we define a dispositional trust score that can be assigned to each study participant.
The dispositional trust score is the sum of the numerical values of the responses to each of the Likert-scale questions (each scored 1--5) that a participant provides on the dispositional trust survey---noting that we invert the scores for any inverted (negative) questions (thus, a ``Strongly Disagree'' becomes a 5 and a ``Strongly Agree'' becomes a 1).
We similarly define and compute a latent trust score based on participants' responses to the pre- and post-study survey instruments (Appendix~\ref{sec:tos-survey}) for each study session.
One consequence of defining these trust scores is that latent trust scores (or changes thereof) can be calibrated or normalized based on participants' dispositional trust scores; e.g., it should perhaps be considered less surprising that an individual with very high dispositional trust achieves a high latent trust score or experiences a large increase in latent trust.
%See Section~\ref{sec:results} for details.

We refer the reader to our supplementary material for complete copies of our survey instruments.

\section{User Study}
\label{sec:user-study}

\begin{table}[!t]
 \caption{Groups for Our User Study.}
 \label{tab:study-groups}
 \begin{tabular}{cccc}
   \toprule
   \# & First Video & Second Video & Final Participant Count \\
   \midrule
   1 & AV Negative & Power Plant (PP) & 20 \\
   2 & AV Negative & Robotic Assembly Line (RAL) & 17 \\
   3 & AV Positive & Power Plant (PP) & 17 \\
   4 & AV Positive & Robotic Assembly Line (RAL) & 21 \\
 \bottomrule
\end{tabular}
\end{table}

\begin{figure}[!t]
  \centering
  \includegraphics[width=\linewidth]{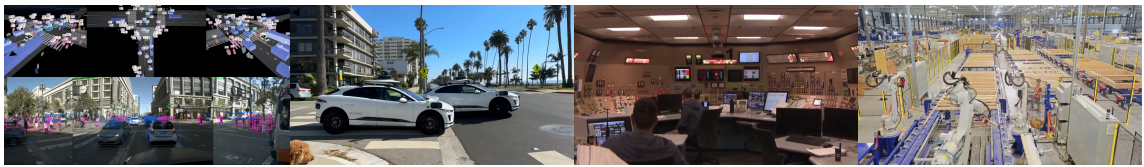}
  \caption{Screenshots from each of the videos different participants groups were shown during our study.  From left to right: AV Positive, AV Negative, power plants (PP), and robotic assembly lines (RAL).}
  \Description{A 1x4 grid of screenshots extracted from the videos used in our study.}
  \label{fig:4x1}
\end{figure}

We conduct a user study using our virtual laboratory to measure the dynamics of human trust in the context of three distinct CPS: autonomous vehicles (AV), power plants (PP), and robotic assembly lines (RAL).
Leveraging existing real-world footage, we assemble 30-minute videos for each CPS (see Figure~\ref{fig:4x1}).
These videos generally show the CPS performing well, as expected, and/or in a trustworthy way; our hypothesis is that passive observation of these videos will engender increased trust in the systems from participants.
The exception is AVs, where we assemble two videos: one where AVs perform well (AV Positive), and one where AVs perform poorly (AV Negative).
We similarly hypothesize that participants in the AV Negative group will experience negative trust calibration over the course of the first session, and we further hypothesize that such participants will have lower trust scores than those in the AV Positive groups on the CPS shown during the \textit{second} session (i.e., learned trust may transfer between CPS or may bias users' general trust toward CPS in advance of the second session).
Each video is generally stripped of original audio (except where necessary to understand a video segment), and an identical, neutral, instrumental soundtrack is added to each video.
Appendix~\ref{sec:videos} provides links to each video.

We divide participants in our study into four groups, illustrated in Table~\ref{tab:study-groups}.
Each group watches either the AV Negative or AV Positive video during the first study session.
The second study session, which is timed one week after the first (to minimize participants' memorization of the survey questions, and to enable participants to potentially better internalize trust learned after the first session), sees participants watch either the PP or RAL videos.
Despite dropout, each group ended up with approximately 20 participants.

We initially recruited 83 participants who attended a first session.
Participants were recruited from student populations at Vanderbilt University and surrounding colleges and universities, as well as through social media advertisements and flyers for the study.
Our only exclusion criteria for recruitment were ensuring participants were adults and that they were eligible to receive compensation.
A \$30 gift card was offered as compensation for completing two study sessions spaced one week apart.
After the first session, the study suffered minor participant dropout; 75 participants completed both sessions.
Nonetheless, pre- and post-survey data from participants who completed only the first session was still able to be used for certain analyses in Section~\ref{sec:results}.
We note that our study was reviewed by our university's institutional review board and was approved as study \#STUDY00000139.

\begin{wrapfigure}{r}{0.4\textwidth}
    \centering
    \includegraphics[width=0.35\textwidth]{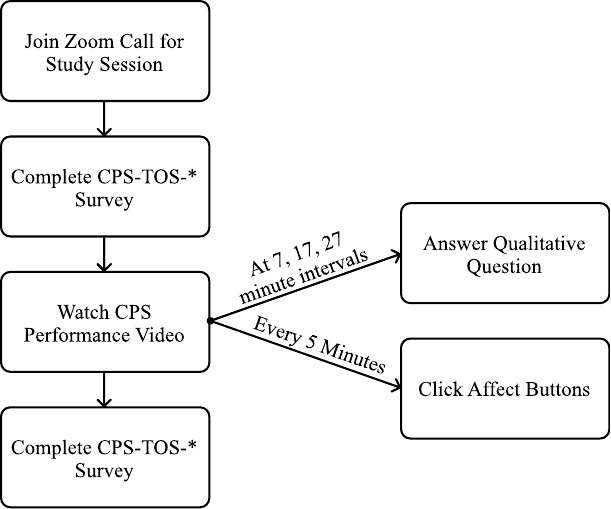}
    \caption{Flowchart of a user study session. Participants complete pre- and post-surveys before and after observing a CPS, during which time they are prompted to provide situational trust data at regular intervals.}
    \Description{The flowchart shows how every 5 minutes in the study, users are prompted to click affect buttons, and thrice during the study, they are prompted with a qualitative question with freeform text response.}
    \label{fig:study-flowchart}
\end{wrapfigure}

Figure~\ref{fig:study-flowchart} illustrates the process of running or participating in a study session.
First, users in a group all join a single Zoom meeting, which is used to speak to and coordinate participants (including providing technical support during the study).
Users then complete a pre-survey---which, in the case of the first session (in all groups), included our dispositional instrument.
The figure labels these surveys CPS-TOS-* to denote ``Cyber-Physical System Trust Observation Sscale'' survey, where the asterisk represents either AV, RAL, or PP.
We replace the words ``the system'' in our pre- and post-surveys with ``autonomous vehicles,'' ``robotic assembly lines,'' or ''power plants'' depending on the session and group as appropriate.

After completing the pre-survey, users log in to our virtual laboratory environment using their email address.
They are then able to click to join the live study session to which they were invited.
Once all users in a group have joined the session, the study administrator starts playing the video in question.
The video plays in a synchronized fashion across all users in the session; users are unable to start or stop the video.
Every 5 minutes, the web application prompts users to click the affect button they most strongly align with based on the last few minutes of video they have witnessed.
%At the 7, 17, and 27 minute markers, qualitative questions are asked of the participant, answers to which they enter via the freeform text chat interface in the application:
At the 7, 17, 27 minute markers, participants were presented with qualitative prompts delivered through the application's freeform text chat. Each prompt was adapted from \citet{lee2004trust}'s performance, process, purpose framework to capture situational trust:
\begin{enumerate}
    \item 7 minutes (Performance): From your perspective, how effectively does the system handle its intended tasks?
    \item 17 minutes (Process): How easy is it for you to understand how the system's operating process works in a situation?
    \item 27 minutes (Purpose): To what extent do you believe the system is designed with people's safety and best interest in mind?
\end{enumerate}
These questions are also customized so that ``the system'' is replaced with the CPS at hand.
Each time they are prompted, users are given 30 seconds to click an affect button, or 40 seconds to answer a qualitative question. To preserve data integrity, affect buttons were disabled after the 30-second prompt window. For qualitative questions, participants were allotted 40 seconds but encouraged to continue typing beyond this period to allow adequate reflection on their observations. The free-form text chat remained open for the duration of the session so participants could complete responses at their own pace.
Notably, users are able to see each other's answers in the chat (an aspect we discuss further in Section~\ref{sec:conclusions}).

After the video concludes, participants are asked to take a post-survey (which, as discussed in Section~\ref{sec:ti}, is identical to the pre-survey).
They are then dismissed from the session.
An entire study session takes approximately 45 minutes to complete.
The study administrator reads a pre-written script for the study session (see supplementary material) to ensure timeliness and consistency between sessions and groups.

Between the two study sessions, we also asked users to complete a demographic survey (see supplementary material).

Although all of the final 75 study participants completed two sessions each and all surveys, there were instances where certain participants did not click an affect button for a particular prompt, or failed to answer a qualitative question during a session.
Despite these minor limitations, our dataset is sufficiently complete to draw various statistically valid conclusions.

\subsection{Results}
\label{sec:results}

\begin{figure}[!b]
  \centering
  \includegraphics[width=\linewidth]{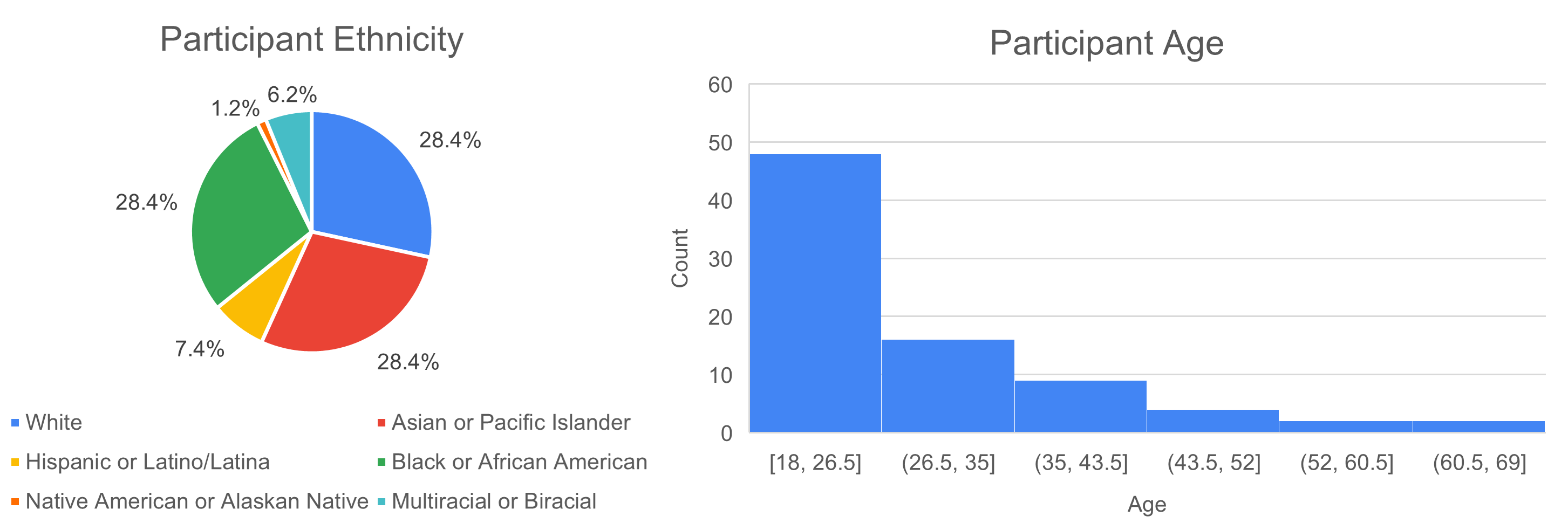}
  \caption{Relevant demographic factors of participants in our study, including ethnicity and age.}
  \Description{Age is heavily skewed towards 18--26 years old, while ethnicity shows exact balance between ``White,'' ``Asian or Pacific Islander,'' and ``Black or African American.''}
  \label{fig:demographics}
\end{figure}

\paragraph{Demographics}
Of the recruited participants, approximately 70\%  were university students, while the remaining 30\% were members of the general community.
Student participants represented a diverse range of academic backgrounds, spanning STEM, social sciences, professional programs, and the arts/humanities.
Approximately 50\% of all study participants fell within an age range of 18--26.
It is known that older adults report greater distrust of technology, and while distrust may not correlate with adoption \cite{knowles2018older}, our population's skew towards younger ages may bias our results towards higher trust scores.
In contrast, participants were somewhat well-balanced when considering ethnicity, with 23 participants reporting ``White,'' 23 participants reporting ``Asian or Pacific Islander,'' and 23 participants reporting ``Black or African American'' (with the remaining participants selecting ``Hispanic or Latino/Latina'' (6), ``Multiracial or Biracial'' (5), or ``Native American or Alaskan Native'' (1)).
Aggregate plots are shown in Figure~\ref{fig:demographics}.

\begin{figure}[!t]
  \centering
  \includegraphics[width=\linewidth]{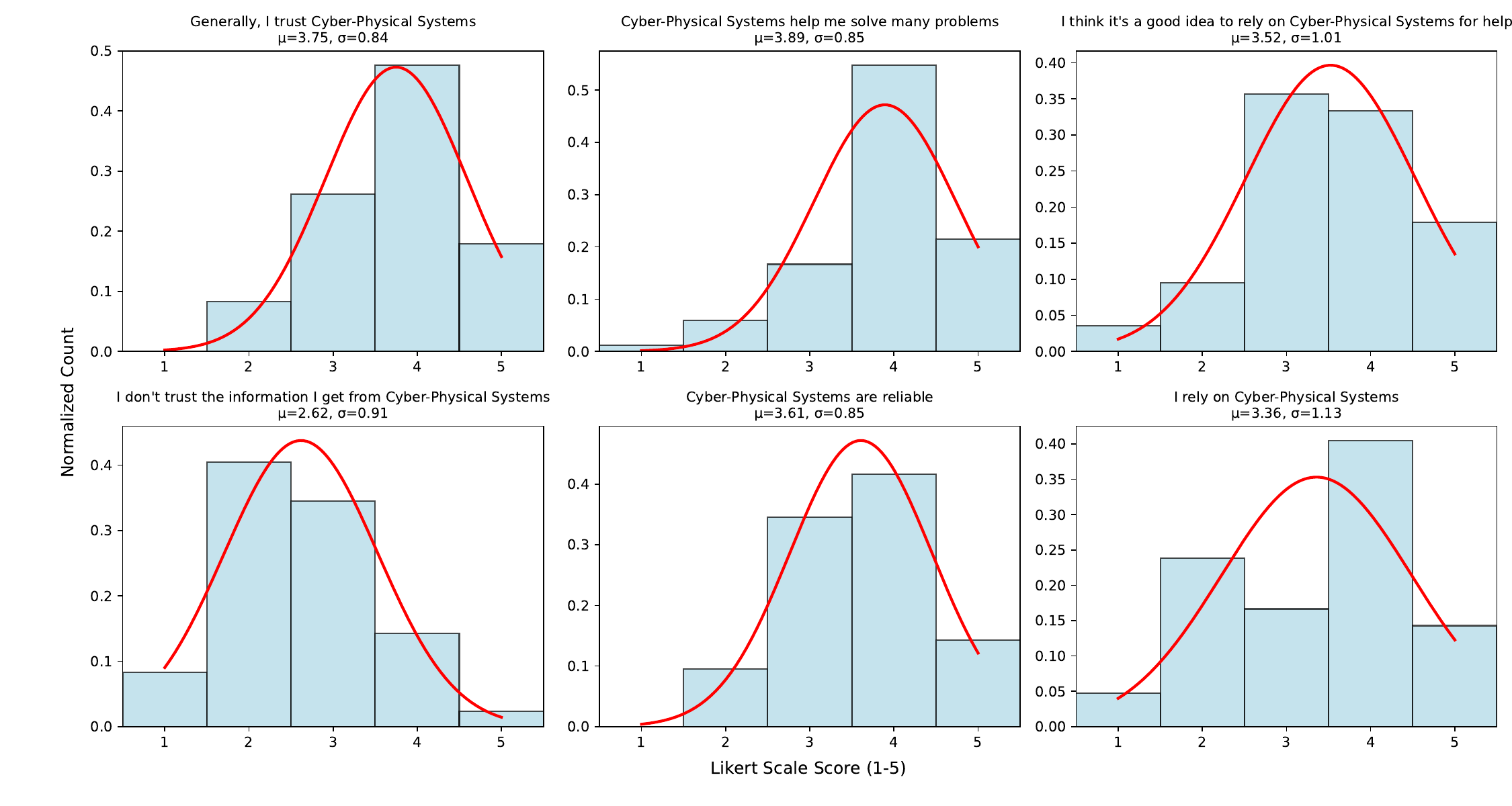}
  \caption{Distributions of dispositional trust factor scores.  Normal distributions are fit to the responses for each item.  Study participants felt neutral or positive towards CPS on all items---though not overly positive, as the highest mean measured was 3.89 (for ``CPS help me solve many problems'').  The bottom-left subfigure represents an inverted question, where 1 indicates greatest trust.}
  \Description{Histograms of data are shown in light blue, and best-fit normal distributions are overlaid in red on each subfigure.}
  \label{fig:dispositional}
\end{figure}

\paragraph{Dispositional Trust}
The first trust measure we analyze is the dispositional trust of study participants towards CPS (see Section~\ref{sec:ti}).
Since this is a brief six-item instrument, we plot the distributions of responses to each item in Figure~\ref{fig:dispositional}.
We note that participants were provided a definition of CPS in the study information sheet (see supplementary material) before the study began: ``In this study, Cyber-Physical Systems (CPS) are engineered systems that combine physical (hardware) and computational (software or cyber) components to
monitor and control real-world processes. Examples include smart home systems, autonomous vehicles, industrial robots, automated safety systems, etc.''

The figure demonstrates that participants' predisposed attitudes towards CPS were generally neutral or slightly positive.
A normal distribution is fit to each histogram (shown as red curves).
Histograms are normalized in each subfigure so that the sum of histogram bars equals one in each subfigure; normal distributions are fit to the normalized histograms.

While Figure~\ref{fig:dispositional} shows results for all groups, we also investigated whether there were any statistically significant differences between groups (see Table~\ref{tab:study-groups}).
For each of the six items in the survey instrument, we performed nonparametric Kruskal-Wallis tests across the four groups, and we also performed pairwise two-sample Kolmogorov-Smirnov (KS) tests for each pair of groups.
\textit{All of these tests yielded statistically insignificant values}, even after considering the Holm–Bonferroni correction for the KS tests.
Thus, we conclude that there were no significant differences in dispositional trust between the groups, which is expected since participants were assigned to groups randomly.

\begin{figure}[!t]
  \centering
  \includegraphics[width=0.5\linewidth]{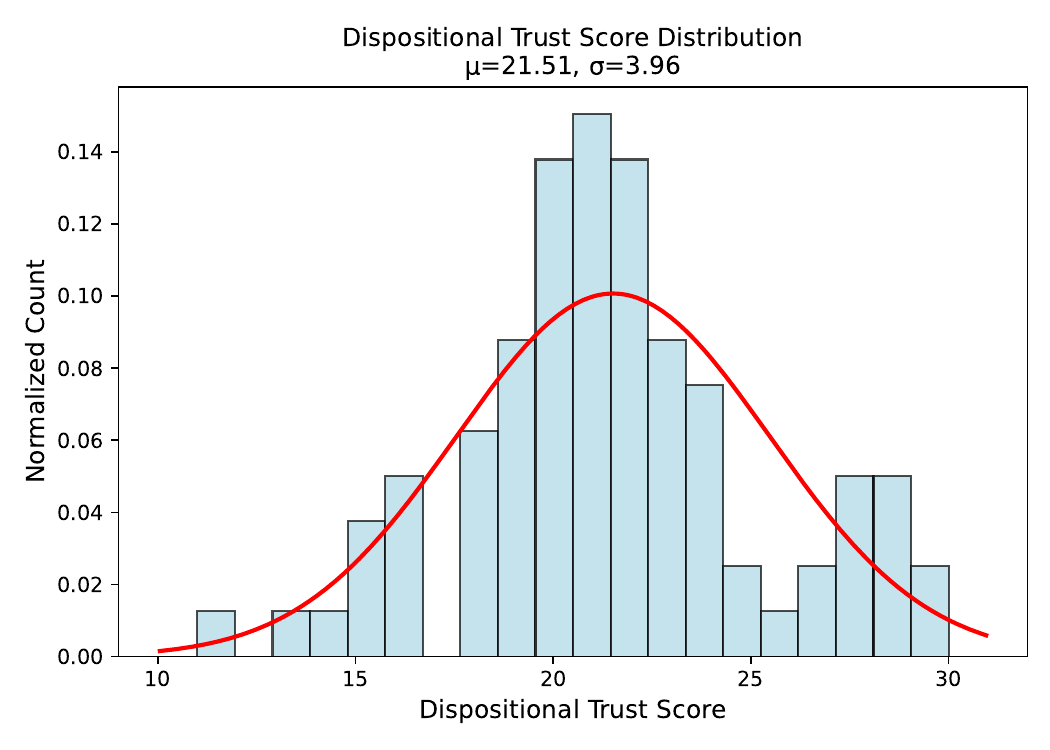}
  \caption{Distribution of dispositional trust scores for all participants.  A normal distribution (red) is fit to the data.  A Shapiro-Wilk test affirms normality of the data.  In our study, scores ranged from 11 to 30, with 30 being the maximum possible score.}
  \Description{A histogram of the data is shown in light blue, and a best-fit normal distribution is overlaid in red.}
  \label{fig:disp-scores}
\end{figure}

As discussed in Section~\ref{sec:ti}, we also compute an aggregate dispositional trust score for each participant.
Since there are no significant differences between groups, we produce one histogram of dispositional trust scores for all participants in the study; see Figure~\ref{fig:disp-scores}.
A Shapiro-Wilk test on the data reveals a score of $W = 0.969$ with $p = 0.04$, implying that participants' dispositional trust scores are normally distributed (as the figure generally suggests).
Therefore, our dispositional trust score may be a reliable single score of individuals' dispositional trust in CPS.

\paragraph{Learned Trust Dynamics in First Session}
The first session of our study saw two of our participant groups watch the AV Positive experiment---30 minutes of footage of an autonomous vehicle behaving well---and the other two groups watch the AV Negative experiment, where various undesirable behaviors occurred with AVs (such as changing into the wrong lane or turning into traffic).

\begin{figure}[!th]
  \centering
  \includegraphics[width=0.9\linewidth]{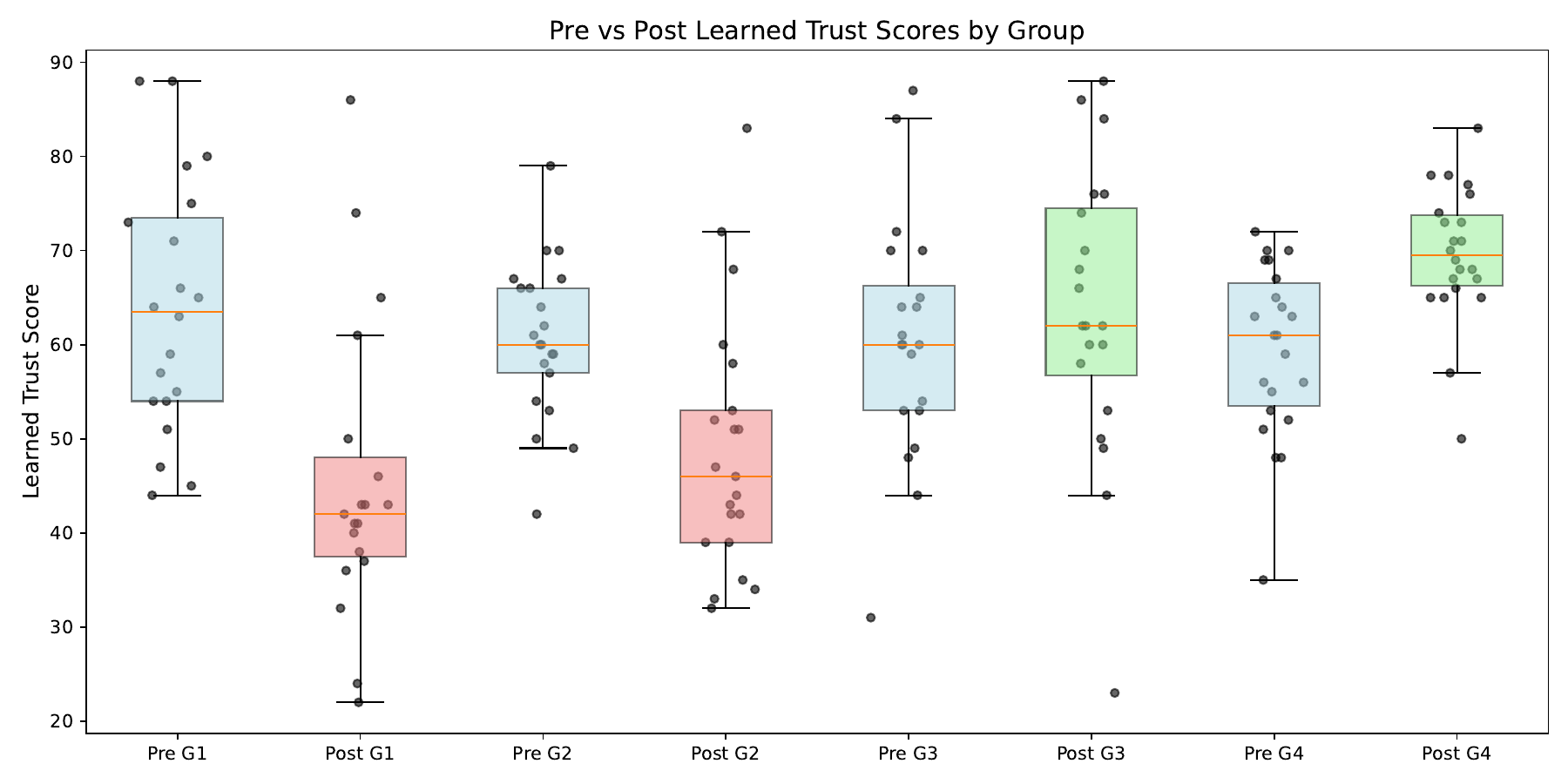}
  \caption{Learned trust scores, as measured by pre- and post-survey instruments, for each study group.  In general, the AV Negative intervention (Groups 1 and 2) appears to lower trust values compared to pre-study values, whereas the AV Positive intervention (Groups 3 and 4) appears to generally increase trust values.}
  \Description{Eight boxplots: one for each group's pre-survey data, and one for each group's post-survey data.  Errorbars/whiskers are included for each box, and raw data points are also scatter-plotted.}
  \label{fig:prepost_all_groups}
\end{figure}

\begin{figure}[ht]
    \centering
    \begin{subfigure}{0.49\linewidth}
        \centering
        \includegraphics[width=\linewidth]{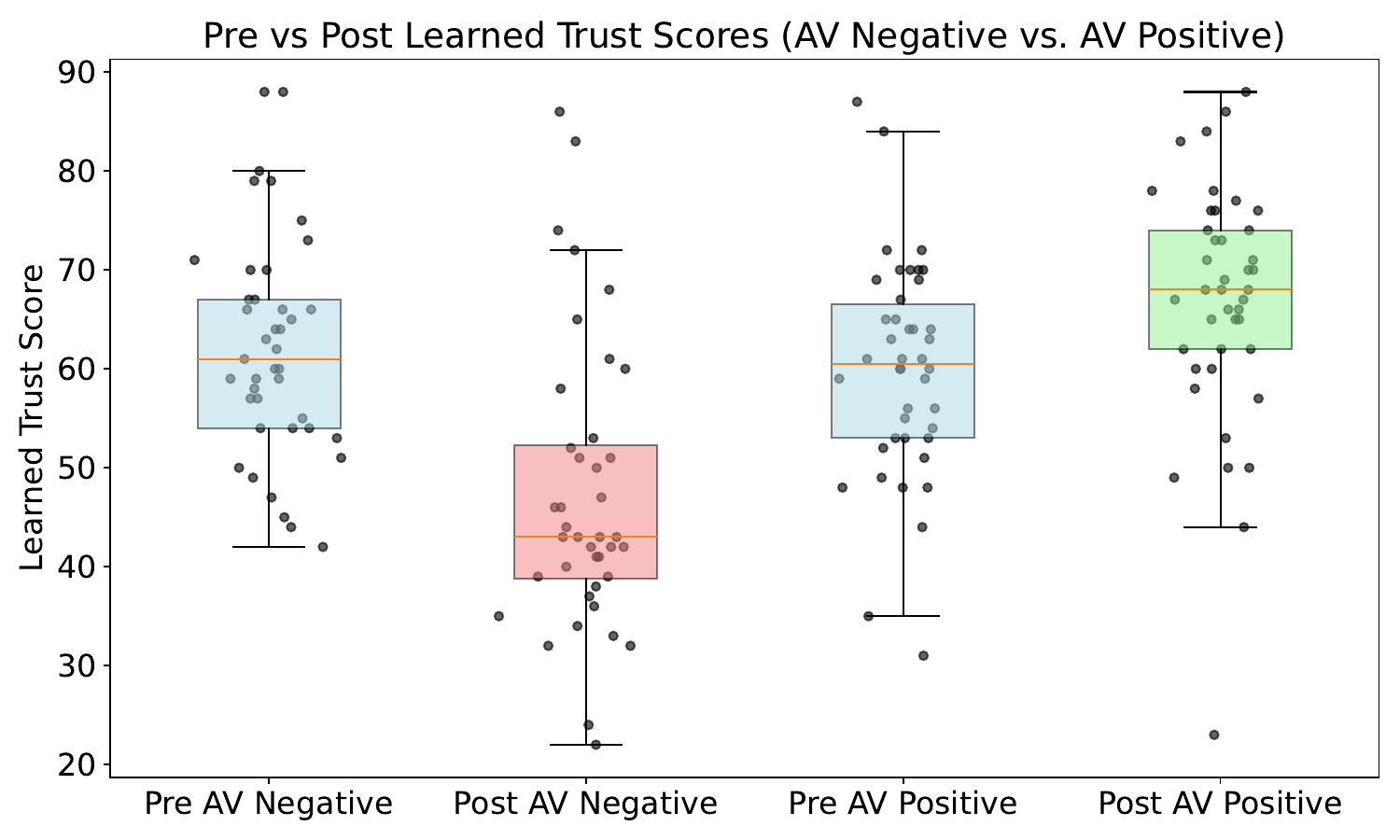}
        \caption{Learned trust scores (no normalization) for the AV Negative and AV Positive groups.}
        \Description{Four box plots: one for the pre-AV Negative condition, one for post-AV Negative,  one for pre-AV Positive, and one for post-AV Positive.  The post-AV Negative box plot is lower than pre-AV Negative, and the post-AV Positive box is higher than the pre-AV Positive box.}
        %\label{fig:sub1}
    \end{subfigure}
    \hfill
    \begin{subfigure}{0.49\linewidth}
        \centering
        \includegraphics[width=\linewidth]{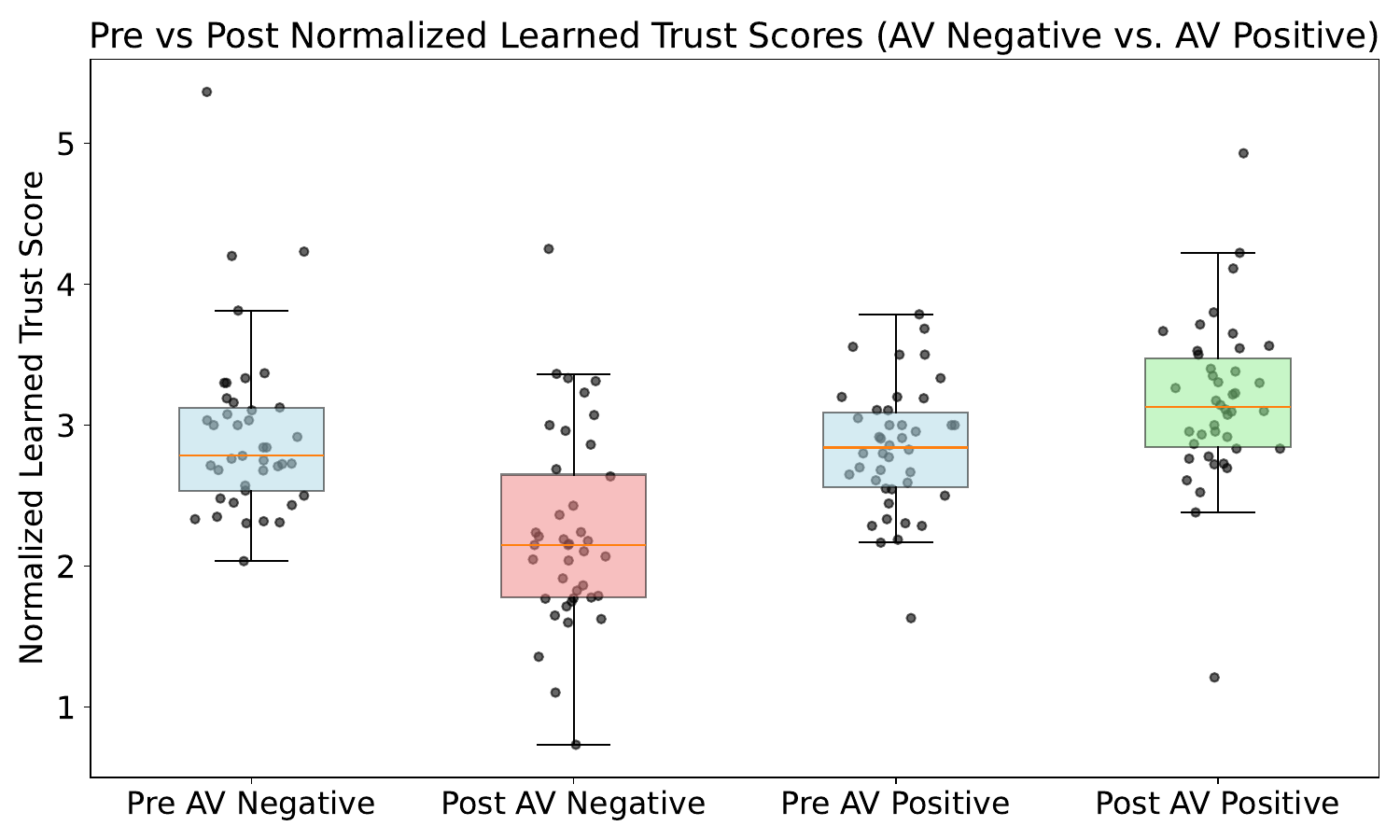}
        \caption{Learned trust scores normalized by participants' dispositional trust scores.}
        \Description{Similar to the subfigure on the left, four boxplots are shown.  The AV Negative intervention appears to generally lower trust, while the AV Positive intervention appears to generally increase trust.}
        %\label{fig:sub2}
    \end{subfigure}
    
    \caption{Learned trust scores, assessed pre- and post-study, for participants who experienced the AV Negative intervention (Groups 1 and 2) and those who experienced the AV Positive intervention (Groups 3 and 4).}
    \label{fig:prepost_combined}
\end{figure}

Analysis of pre- and post-survey data confirms our hypothesis that individuals' learned trust is statistically significantly affected by watching either AV Positive or AV Negative.
To measure the effect of the intervention, we computed a pre- and post- composite learned trust score for each participant, analogous to how the dispositional trust was computed.
Figure~\ref{fig:prepost_all_groups} displays box plots for the pre- and post- learned trust scores for each group.
Raw data points are scattered in gray.
Blue boxes represent pre-session conditions, red boxes represent post-session conditions for AV Negative groups, and green boxes represent post-session conditions for AV positive groups.
Wilcoxon signed rank tests confirm statistically significant differences between pre- and post-conditions for Group 1 ($W = 9.0$, $p < 0.001$\footnote{In this group, one participant completed the pre-survey but not the post-survey.  In order to apply the Wilcoxon test, we removed their sample from the pre-survey data, even though that data point is included in Figure~\ref{fig:prepost_all_groups} and Figure~\ref{fig:prepost_combined}.}), Group 2 ($W = 14.5$, $p < 0.001$), and Group 4 ($W = 1.5$, $p < 0.001$).
The exception is Group 3, where we compute $W = 57.5$, $p = 0.131$, suggesting an insignificant difference.
However, we remark that there was an unexpected outlier in the Group 3 post-survey (as is evident in Figure~\ref{fig:prepost_all_groups}), where the participant indicated the lowest level of trust for almost all survey items, despite being exposes to positive AV behavior.
Removing that participant from the pre- and post- score data changes the Wilcoxon test results for Group 3 to $W = 44.5$, $p = 0.074$, which is on the borderline of statistical significance.

We note that in the first study session, Groups 1 and 2 undergo the same intervention (AV Negative), and Groups 3 and 4 both experience AV Positive.
Thus, we also analyze the two group pairs combined; see Figure~\ref{fig:prepost_combined} Left.
In this case, the outlier effects of the participant in Group 3 mentioned above are greatly ameliorated; even when leaving in that participant, Wilcoxon tests report statistically significant changes in learned trust score for the AV Negative groups ($W = 41.5$, $p < 0.001$) and AV Positive groups ($W = 99.0$, $p < 0.001$).

Finally, although it is possible that learned trust scores already include the biases of participants' varying dispositional trust, we assume for sake of argument that they do not.
In such a scenario, it is useful to consider how participants' learned trust scores evolve over the first session when normalized by their initial dispositional trust score.
To assess this, we simply divide each participant's pre- and post- learned trust scores by their dispositional trust score and re-run the statistical analyses performed above.
Figure~\ref{fig:prepost_combined} Right shows the result of this normalization, when considering AV Positive and AV Negative groups together.
The Wilcoxon test outcomes are $W = 38.0$, $p < 0.001$ for AV Negative, and $W = 111.5$, $p < 0.001$ for AV Positive.
Thus, whether or not normalization is applied---i.e., whether or not learned trust includes the effects of dispositional trust in this experiment---we conclude that there are statistically significant decreases (increases) in learned trust for the AV Negative (Positive) groups by virtue of the study intervention.

\begin{figure}[!t]
  \centering
  \includegraphics[width=0.9\linewidth]{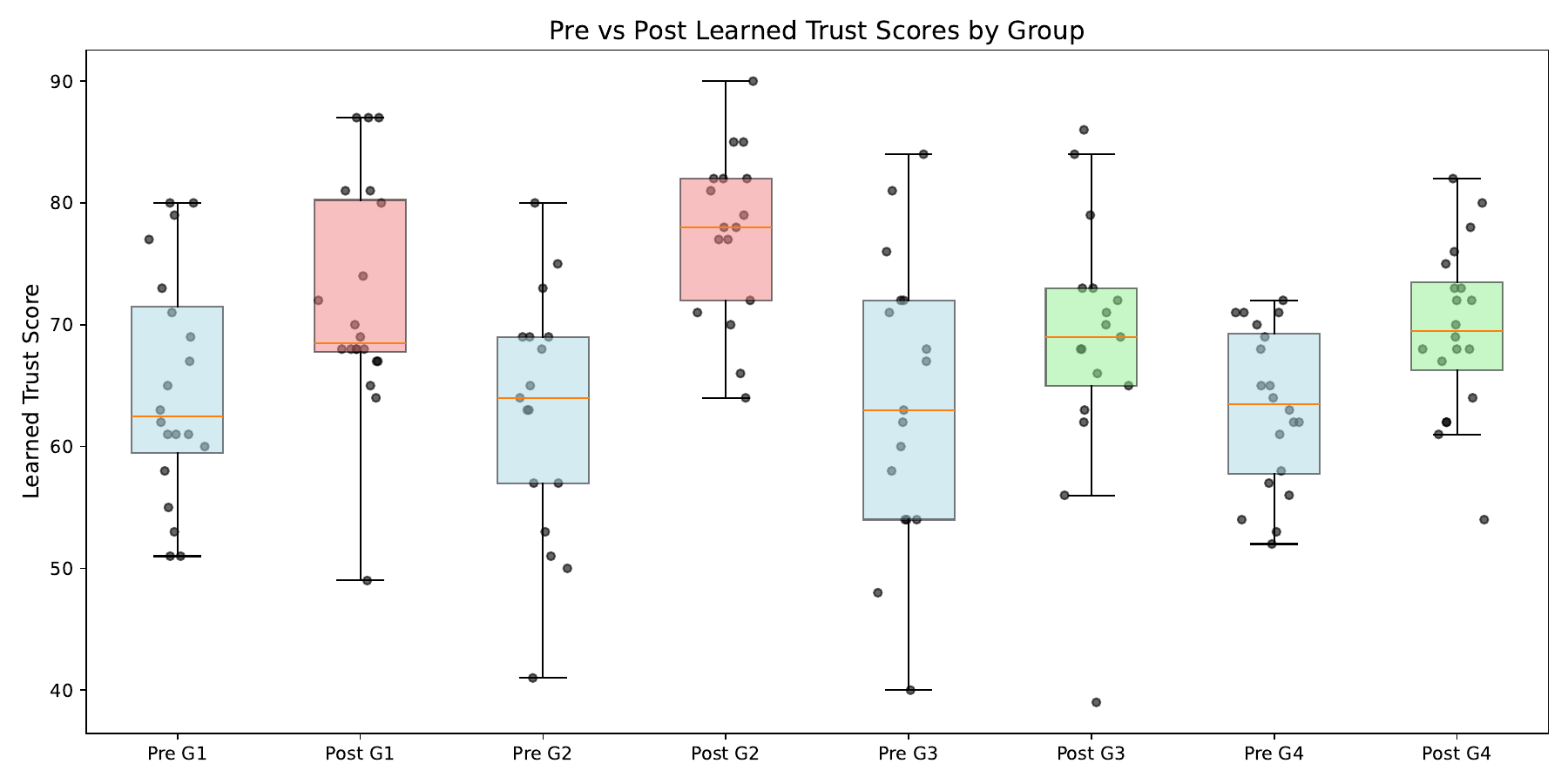}
  \caption{Learned trust scores, as measured by pre- and post-survey instruments, for each study group, in the second session of the study.  Trust appears to increase for every study group.  Generally, higher trust scores are observed for Groups 1 and 2, which were first exposed to AV Negative, versus Groups 3 and 4, which were first exposed to AV Positive.}
  \Description{Eight boxplots: one for each group's pre-survey data, and one for each group's post-survey data.  Errorbars/whiskers are included for each box, and raw data points are also scatter-plotted.}
  \label{fig:second_session_all_groups}
\end{figure}

\paragraph{Learned Trust Dynamics in Second Session}
We perform a similar analysis for the second session.
In the second session, which occurred one week after the first, participant groups were exposed to either the power plants (PP) or robotic assembly lines (RAL) video (see Table~\ref{tab:study-groups}); thus, with four distinct conditions, we do not group AV Negative and AV Positive as in Figure~\ref{fig:prepost_combined}.
Instead, analogously to Figure~\ref{fig:prepost_all_groups}, Figure~\ref{fig:second_session_all_groups} displays the pre- and post-survey results for each of the four groups.
All four groups are equivalent at the start of the second session (no statistically significant differences were observed), which was slightly surprising given groups' varied post-survey learned trust scores after the first session.
However, we note that the pre- and post-surveys for the second session used the words ``power plants'' or ``robotic assembly lines'' rather than ``autonomous vehicles,'' so we conjecture that participants did not transfer their learned trust of one CPS to another, distinct CPS.

Yet important differences emerged when considering the post-survey data.
\textit{All} groups had statistically significantly higher trust values in post-survey data compared to pre-survey data, per Wilcoxon signed rank tests ($p < 0.001$).
But as evidenced in the figure, Groups 1 and 2 (which were first exposed to AV Negative) both had higher trust scores, on average, than Groups 3 and 4 (which were first exposed to AV positive).
Two-sided KS tests reveal that the differences between groups 1 and 3 ($W = 0.203$, $p = 0.745$) and 1 and 4 ($W = 0.2$, $p = 0.832$) were not statistically significant, although the differences between groups 2 and 3 ($W = 0.529$, $p = 0.016$) and 2 and 4 ($W = 0.556$, $p = 0.004$) were.
%The potential difference in significance levels could be explained by differences between the PP and RAL videos themselves
%This is not explained by potential differences between PP and RAL, since Groups 1 and 3 watched PP and Groups 2 and 4 watched RAL.
%The PP and RAL videos both display CPS operating positively, so there is not a substantial difference between the videos for purposes of our study.
%Rather, 
Although only half of these comparisons allow us to definitively establish significant differences, our statistical results in conjunction with the visual evidence of Figure~\ref{fig:second_session_all_groups} suggest that priming participants with experiencing a negative CPS led them to place even greater trust in other CPS that behaved positively.
A future study with additional participants could help reduce uncertainty and further validate this hypothesis.

\paragraph{Situational Trust Dynamics: Affect Buttons}
As discussed in Sections~\ref{sec:design} and \ref{sec:ti}, we also prompt users to click affect buttons and answer qualitative questions (with freeform text responses) during each study session.
These measurements are designed to assess situational trust---how individuals' trust is affected in real-time as they passively observe CPS behavior.

\begin{figure}[!t]
  \centering
  \includegraphics[width=0.7\linewidth]{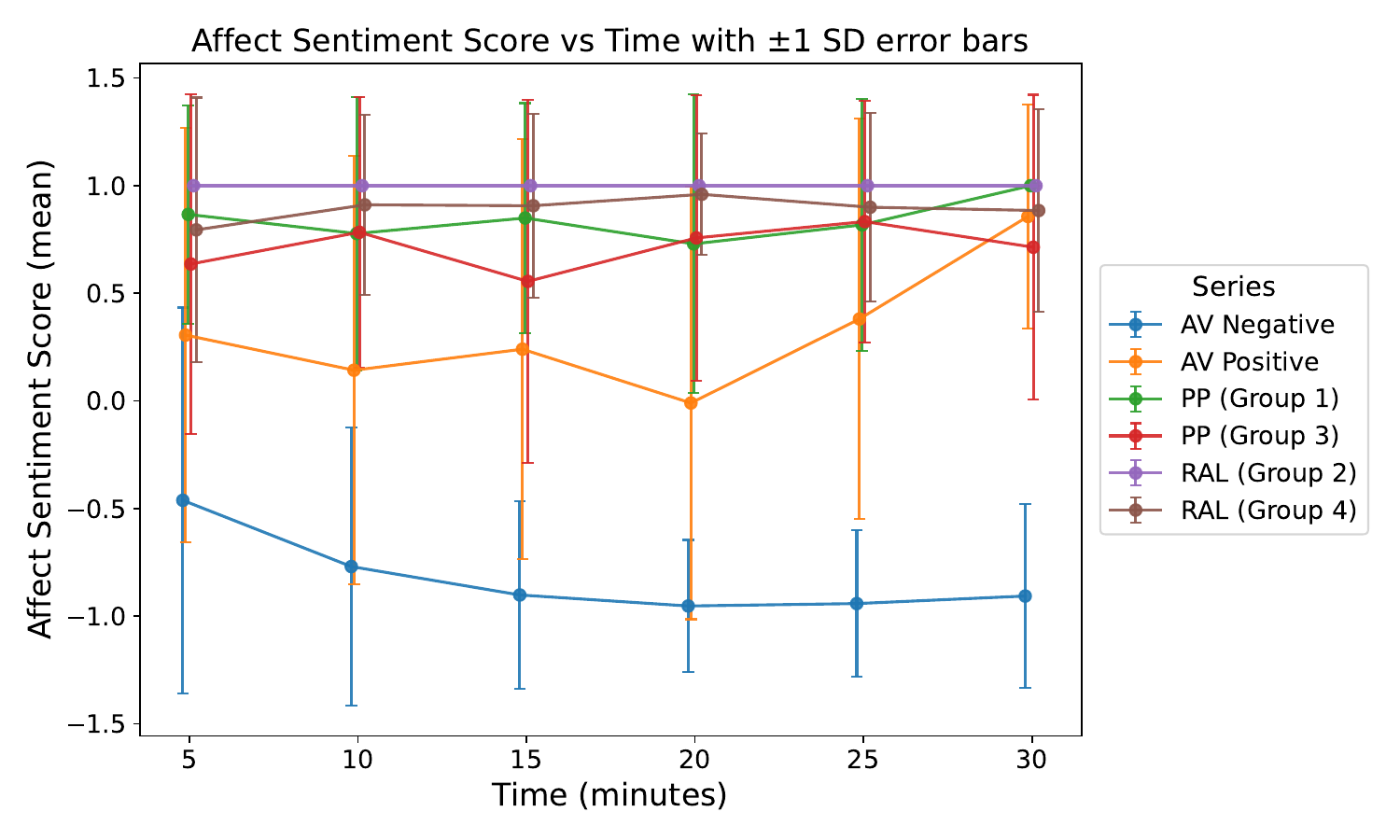}
  \caption{Distributions of the affect sentiment score over time during both the first and second sessions of our user study. Error bars indicate $\pm$1 standard deviation from the mean (circles).  Due to the identical interventions applied to Groups 1 and 2, and to Groups 3 and 4, in the first study session, there are only two lines shown for the first session (AV Negative and AV Positive).  One line per group is shown for the second session.  In the second session, Group 2 demonstrates strongly positive affect sentiment scores, while in contrast, AV Negative participants during the first session record strongly negative scores.}
  \Description{Six lines are shown: two for the first session (combining AV Negative and AV Positive groups), and four for the second session.  Each line is constructed with a point at each five-minute marker representing the mean affect sentiment score at that time, and vertical errors bars around each point indicate plus or minus one standard deviation.}
  \label{fig:realtime_affect}
\end{figure}

We first evaluate the affect button responses.
Each affect button has a polar opposite button (see Figure~\ref{fig:affect-buttons}); we assign a score of 1 to each positive button and a score of -1 to each negative button.
This can be viewed as a binary sentiment score.
Figure~\ref{fig:realtime_affect} plots the mean and standard deviation of this sentiment score over time.
We note that subjects were intended to click the affect buttons at five-minute intervals throughout the study (the session automatically prompted users to do so at the appropriate times).
A few participants clicked the affect buttons unprompted; these data points were removed in the figure.
Also, at each five-minute mark, the data points in the figure are horizontally offset from each other by 0.2, for ease of readability.

The plot illustrates that the AV Negative group quickly reaches a low sentiment score, which stays distributed quite near -1 throughout the study.
In contrast, the AV Positive group's affect sentiment score tends to increase throughout the session to a distribution centered almost at +1.
Due to the study's limited sample size, and perhaps more importantly, due to the binary nature of this metric, the one-standard-deviation error bars on each data point are relatively large, which inhibits making conclusions about statistical significance.
Another potential contributor may be recency bias; participants may have weighted their judgments more heavily on what they observed in the final moments leading up to each prompt, rather than integrating their observations across the entire preceding interval.
While there is not sufficient evidence for this, we acknowledge it as a possible source of variability that may have contributed to the inflated error bars.
Nonetheless, it is clear, for instance, that the AV Positive and AV Negative groups, while starting within 1SD of each other, end up well-separated by more than 1SD.

Figure~\ref{fig:realtime_affect} also includes data from the second study session, where the four subject groups had differing results.
Although there are visible differences, the distributions at each time overlap among all groups (again, due in part to the binary nature of the sentiment score used here), so it is difficult to conclude that there are statistically significant differences in this situational trust metric alone.
We remark that while affect buttons do indicate positive or negative sentiment, they do not measure how strongly an emotion is felt.

\paragraph{Situational Trust Dynamics: Text Responses}
As mentioned in Section~\ref{sec:user-study}, we prompt users with qualitative trust questions during the study after 7, 17, and 27 minutes.
To quantitatively assess this data, we also considered a sentiment score based approach.
With freeform text data, we can produce continuous sentiment score values from -1 to +1.
To compute such scores, we use a transformer-based approach, leveraging the DistilBERT \cite{sanh2019distilbert} model fine-tuned on the SST-2 dataset \cite{socher-etal-2013-recursive}, which is available on HuggingFace as \texttt{distilbert/distilbert-base-uncased-finetuned-sst-2-english}.
The neural network produces probability scores for the text to have positive or negative sentiment; we compute a continuous sentiment score as $s = P(\text{positive}) - P(\text{negative})$.

\begin{figure}[!t]
  \centering
  \includegraphics[width=0.7\linewidth]{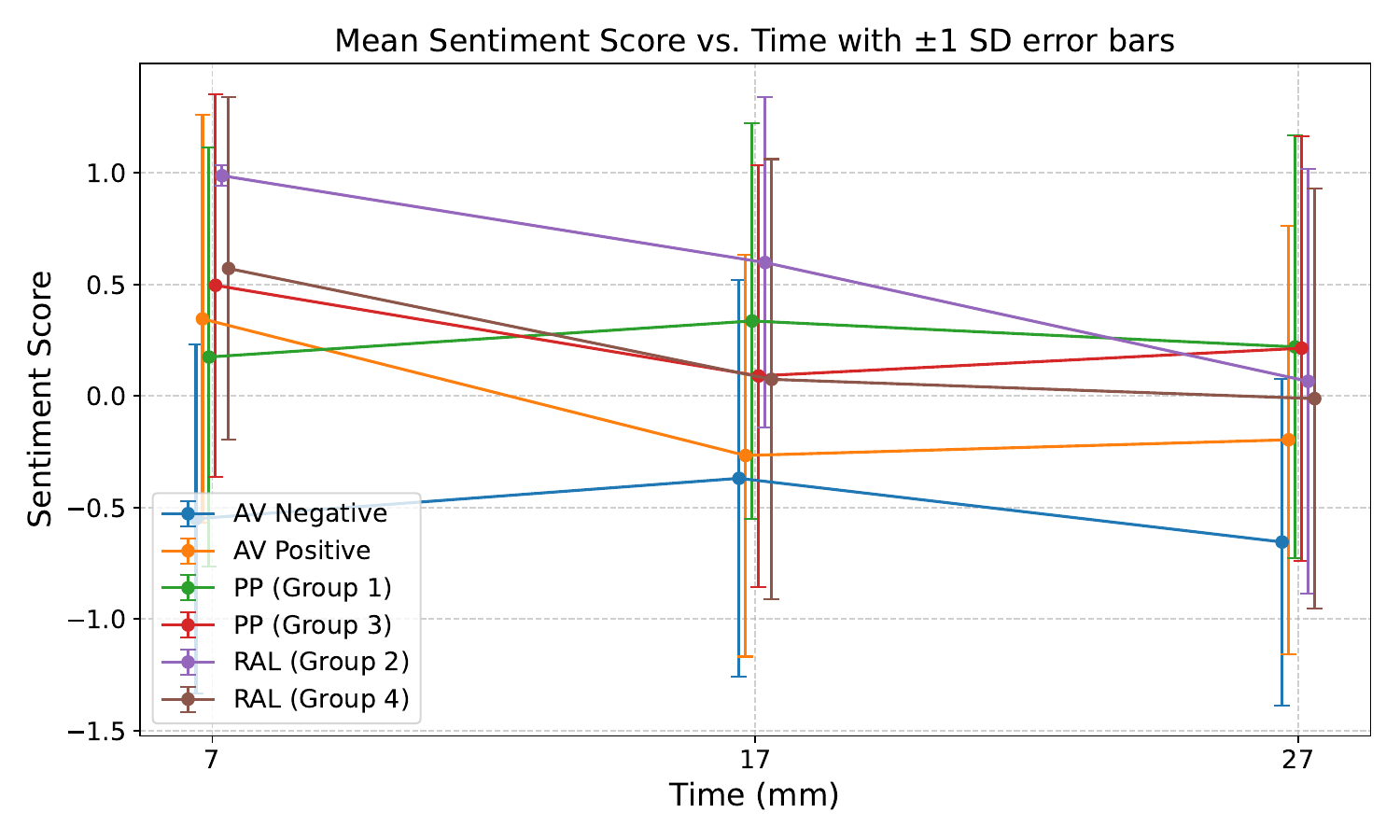}
  \caption{Distributions of the sentiment score over time during both the first and second sessions of our user study, based on subjects' answers to our qualitative trust questions. Error bars indicate $\pm$1 standard deviation from the mean (circles).  Due to the identical interventions applied to Groups 1 and 2, and to Groups 3 and 4, in the first study session, there are only two lines shown for the first session (AV Negative and AV Positive).  One line per group is shown for the second session.}
  \Description{Six lines are shown: two for the first session (combining AV Negative and AV Positive groups), and four for the second session.  Each line is constructed with points at 7, 17, and 27 minutes, representing the mean sentiment score at that time, and vertical errors bars around each point indicate plus or minus one standard deviation.}
  \label{fig:realtime_qual}
\end{figure}

Figure~\ref{fig:realtime_qual} shows the results of this analysis and is formatted similarly as Figure~\ref{fig:realtime_affect} (including filtering out responses of users not provided at the times corresponding to when questions were asked).
Even with a continuous sentiment score, there are relatively large error bars, although we observe that AV Negative---the only experimental condition where participants were exposed to a CPS performing poorly---consistently records the lowest mean values.

To check for any significant differences in the results, we performed Mann-Whitney U tests between each series.
We found statistically significant differences ($p < 0.001$) between the AV Negative group and \textit{each} other group, as well as between the AV Positive Group and RAL (Group 2) ($p < 0.001$)---with Group 2 being the group that was exposed to AV Negative during the first session.
Thus, there is another piece of evidence suggesting the possibility that prior exposure to a negatively-performing CPS elicits greater situational trust from participants when they are subsequently exposed to a positively-performing CPS.

Overall, from analyzing both affect button and qualitative question responses, and given that AV Negative was the only trust intervention with negative CPS behavior, we can conclude that situational trust was most strongly (and indeed, statistically significantly) affected by whether the trust intervention shows a CPS behaving positively or negatively.
It was not significant \textit{which} CPS was being presented to participants, which agrees with the authors' intuition.
Furthermore, in the second session, situational trust was generally not impacted by which intervention subjects had experienced in the first session.
Possible explanations for this include: that the week between sessions was long enough to allow subjects' trust biases to reset; that participants' mental models distinguished between trust in one type of CPS versus another; and that participants' learned trust from the first session was truly an independent human factor from situational trust.

%\begin{table}
%  \caption{Frequency of Special Characters}
%  \label{tab:freq}
%  \begin{tabular}{ccl}
%    \toprule
%    Non-English or Math&Frequency&Comments\\
%    \midrule
%    \O & 1 in 1,000& For Swedish names\\
%    $\pi$ & 1 in 5& Common in math\\
%    \$ & 4 in 5 & Used in business\\
%    $\Psi^2_1$ & 1 in 40,000& Unexplained usage\\
%  \bottomrule
%\end{tabular}
%\end{table}

% \begin{figure}[h]
%   \centering
%   \includegraphics[width=\linewidth]{sample-franklin}
%   \caption{1907 Franklin Model D roadster. Photograph by Harris \&
%     Ewing, Inc. [Public domain], via Wikimedia
%     Commons. (\url{https://goo.gl/VLCRBB}).}
%   \Description{A woman and a girl in white dresses sit in an open car.}
% \end{figure}

\section{Discussion and Future Work}
\label{sec:conclusions}

Our study found statistically significant evidence that human trust in cyber-physical systems, as measured through our customized trust instruments, can be affected by merely watching videos of CPS, without direct in-person experience.
Our significant results include:
\begin{itemize}
    \item Learned trust in a CPS increased (decreased) when participants were given a trust intervention that involved passive observation of a positively (negatively) behaving CPS---regardless of the CPS being considered.
    \item Participants who experienced the AV Negative intervention in the first session had higher post-survey learned trust values after a positive CPS intervention in the second session (significantly, at least, in the case of Group 2, which was shown the robotic assembly lines video in the second session).
    \item Situational trust did not appear to depend on any prior learned trust, but it was significantly affected based on whether the current intervention being shown to a participant was positive or negative.
    \item Dispositional trust was established to be statistically indistinguishable between groups, suggesting there were no biases in our results due to our (randomized) group assignment procedure.
\end{itemize} %
%Crucially, these results were established using our system in which participants only passively observe videos of CPS; no interaction with or in-person observation of the underlying CPS is required to effect dynamic trust calibrations in participants.
As researchers, companies, and policymakers often seek to promote human trust (or mitigate factors that compromise it), we believe our work has  applications across various domains where CPS are involved.

While our study documents potentially useful findings, we believe that additional interesting work could be immediately conducted using our developed web platform as a virtual laboratory for human trust.
For instance, while our study had all participants watch videos collaboratively in groups, it would be interesting to compare how trust dynamics evolve in the case where each participant watches videos and records their reactions individually, without witnessing the responses of others during the study.
Research suggests that trust is inherently socially propagative: individuals revise their confidence in systems not solely based on direct observations, but also by interpreting others' reactions, attitudes, and behaviors \cite{sherchan2013survey}.
This dynamic is further amplified through homophily---the tendency for individuals to align more closely with those who share similar views---leading participants to be more easily influenced by peers they perceive as like-minded.
%\citet{au2011strength} extend this understanding by providing empirical evidence that trust relationships in social networks mirror existing similarities and actively drive increasing convergence in opinions over time through ongoing social influence.
%Beyond the gradual alignment of peer opinions and trust evaluations, direct social pressure also plays a critical role in shaping trust.
\citet{li2006using} demonstrate that subjective norm---the perceived social pressure from important others---directly and significantly shapes individuals' trusting beliefs, attitudes, and intentions toward information systems.
%These influences, at times, can substantially outweigh individuals' initial personal assessments, suggesting that shifts in trust trajectories may occur even without explicit persuasion \cite{au2011strength,lee2004trust}.
Taken together, these insights underscore the potential value of teasing out the role of peer influence in shaping trust development within collaborative environments.

Furthermore, although our virtual laboratory supports real-time, live video streams or webcam feeds, the user study conducted in this paper used only pre-recorded videos.
Our platform could be used to replicate studies like \citet{bos2001being} or \citet{nguyen2007multiview}, or to conduct new experiments in these contexts.
With the recent rise of online streaming technologies such as Twitch, Instagram Live, and TikTok LIVE, it is worthwhile to conduct further inquiry into how trust shapes and is shaped by these environments and tools.

We also note that the present paper is not a comparative analysis of different trust interventions.
Yet, as future work, it would be quite interesting to measure the difference (if any) in effect size between trust interventions when using a passive observation system (as in the present work), a simulator (either physical \cite{9304627} or virtual), or a real instance of a CPS.
Intuitively, one might expect real-world interactions to induce greater trust, but works such as \citet{bos2001being} illustrate that interactions in virtual space can yield similar trust outcomes as those in real space.

% TODO do we want to try to do any analysis by demographic factor?  (age/race)?
%Lastly, research \cite{10.1109/SMC.2016.7844677, 10.1145/2559206.2559974} has identified that cultural factors can play a significant role in the dynamics of trust.  While the present study assesses certain demographic factors of our participant group, interesting future work could involve assessing human-CPS trust dynamics in different countries and cultures.

Finally, beyond the realm of cyber-physical systems, we believe the rapid proliferation and commoditization of high-quality generative AI video and deepfake tools urgently increases the importance of accurate models and taxonomies for human trust, whereby HCI researchers and practitioners can design, build, and evaluate systems for a world where seeing should no longer be believing \cite{chesney2019deepfakes,kalpokas2020problematising,abonamah2021commoditization,etienne2021future,vaccari2020deepfakes}.  We hope that the community's future work at the intersection of trust and video content will help, in some small part, navigate such historic shifts.

\begin{acks}
We thank Bibek Dhungana, Nurshat Mangnike, and Shihan Cheng for their help in running user study sessions for this paper.
The user study was conducted with approval (18 Aug.\ 2025) from the Vanderbilt University Internal Review Board, IRB \#STUDY00000139.
Zhi Hua Jin was supported in part by a Lanier Family Immersion Award.
\end{acks}

%\clearpage

\bibliographystyle{ACM-Reference-Format}
\bibliography{references}

\clearpage

\appendix

\section{Trust Survey Instrument Details}
\label{sec:tos-survey}

Section~\ref{sec:ti} introduces our construction of a survey instrument for measuring learned trust.
To create this survey, we began with keywords or key phrases identified by \citet{lee2004trust}, and we sought to construct statements that connected each keyword to the process, performance, or purpose (using the language of \citet{lee2004trust}) of a cyber-physical system.
These statements were all presented to study participants with a standard five-point Likert scale rating: Strongly Disagree, Disagree, Neutral, Agree, or Strongly Agree.
For each item, we identified similar items/statements in the trust literature, beyond \citet{lee2004trust}, to properly justify the developed statements.

{
\footnotesize
\rowcolors{2}{white}{gray!15}
\begin{longtable}{cp{0.25\textwidth}p{0.20\textwidth}p{0.45\textwidth}}
   \caption{Items in Our Pre- and Post-Survey Trust Instrument. \label{tab:tos-survey}} \\
   \toprule
   \# & \multicolumn{1}{c}{Item} & \multicolumn{1}{c}{Key Word or Phrase} \cite{lee2004trust} & \multicolumn{1}{c}{Use in Related Works} \\
   \midrule
   1 & The system performs reliably. & Reliable &
   \begin{itemize}[before={\begin{minipage}{\hsize}},after={\end{minipage}}]
    \item ``The system is reliable'' \cite{jian2000foundations}
    \item ``Excel is a very reliable piece of software'' \cite{mcknight2011trust}
    \item ``The system works reliably'' \cite{korber2018theoretical}
    \item ``What \% of the time will this robot be [...] reliable'' \cite{schaefer2016measuring}
    \item ``Please rate the robot [...] reliable'' \cite{malle2021multidimensional}
    \item ``The system performs reliably'' \cite{madsen2000measuring}
   \end{itemize} \\
   2 & The system operates consistently under similar conditions. & Consistent &
   \begin{itemize}[before={\begin{minipage}{\hsize}},after={\end{minipage}}]
    \vspace*{2pt}
    \item ``The system analyzes problems consistently'' \cite{madsen2000measuring}
    \item ``The system performs consistently'' \cite{wojton2020initial}
    \item ``What \% of the time will this robot [...] act consistently'' \cite{schaefer2016measuring}
    \item ``Please rate the robot [...] consistent'' \cite{malle2021multidimensional}
    \item ``The tank-spotting aid's advice consistently helps me perform well'' \cite{chancey2017trust}
   \end{itemize} \\
   3 & A malfunction of the system is likely during normal use. & Functional (``What the automation does'') &
   \begin{itemize}[before={\begin{minipage}{\hsize}},after={\end{minipage}}]
    \vspace*{2pt}
    \item ``Excel does not malfunction for me'' \cite{mcknight2011trust}
    \item ``A system malfunction is likely'' \cite{korber2018theoretical}
    \item ``\% of time robot [...] malfunction'' \cite{schaefer2016measuring}
   \end{itemize} \\
   4 & The system is capable of taking over complicated tasks. & Capable (``characteristics such as [...] ability'') &
   \begin{itemize}[before={\begin{minipage}{\hsize}},after={\end{minipage}}]
    \vspace*{2pt}
    \item ``The system is capable of interpreting situations correctly'' \cite{korber2018theoretical}
    \item ``The system is capable of taking over complicated tasks'' \cite{korber2018theoretical}
    \item ``Please rate the robot [...] capable'' \cite{malle2021multidimensional}
    \item ``I am confident about the system’s capabilities'' \cite{korber2018theoretical}
   \end{itemize} \\
   5 & I can rely on the system to function/operate properly. & Reliable &
   \begin{itemize}[before={\begin{minipage}{\hsize}},after={\end{minipage}}]
    \vspace*{2pt}
    \item ``I can rely on the system to function properly'' \cite{madsen2000measuring}
    \item ``I feel comfortable relying on the information provided by the system'' \cite{wojton2020initial}
    \item ``The tank-spotting aid’s advice reliably helps me perform well'' \cite{chancey2017trust}
   \end{itemize} \\
   6 & The system responds predictably under similar conditions. & Predictable (``Predictability'') &
   \begin{itemize}[before={\begin{minipage}{\hsize}},after={\end{minipage}}]
    \vspace*{2pt}
    \item ``I know it is predictable'' \cite{gefen2003trust}
    \item ``The system reacts unpredictably'' \cite{korber2018theoretical}
    \item ``What \% of the time will this robot be [...] predictable'' \cite{schaefer2016measuring}
    \item ``Please rate the robot [...] predictable'' \cite{malle2021multidimensional}
    \item ``The system responds the same way under the same conditions at different times'' \cite{madsen2000measuring}
    \item ``I am rarely surprised by how the system responds'' \cite{wojton2020initial}
   \end{itemize} \\
   7 & The system is competent at completing tasks. & Competent (``Competency'') &
   \begin{itemize}[before={\begin{minipage}{\hsize}},after={\end{minipage}}]
    \vspace*{2pt}
    \item ``Excel provides competent guidance (as needed) through a help function'' \cite{mcknight2011trust}
    \item ``Please rate the robot [...] competent'' \cite{malle2021multidimensional}
    \item ``What \% of the time will this robot be [...] incompetent'' \cite{schaefer2016measuring}
    \item ``The system has sound knowledge built in'' \cite{madsen2000measuring}
   \end{itemize} \\
   8 & I understand how the system operates. & Understandable (``process [...] describes how the automation operates'') &
   \begin{itemize}[before={\begin{minipage}{\hsize}},after={\end{minipage}}]
    \vspace*{2pt}
    \item ``I was able to understand why things happened'' \cite{korber2018theoretical}
    \item ``I understand how the system will assist me with decisions I have to make'' \cite{madsen2000measuring}
    \item ``I understand how the tank-spotting aid will help me perform well'' \cite{chancey2017trust}
    \item ``I will be able to perform well the next time I use the tank-spotting aid because I understand how it behaves'' \cite{chancey2017trust}
    \item ``I understand how the system executes tasks'' \cite{wojton2020initial}
    \item ``I understand the capabilities of the system'' \cite{wojton2020initial}
   \end{itemize} \\
   9 & It is clear how the system functions in ways that reflect integrity and compliance with standards. & Integrity &
   \begin{itemize}[before={\begin{minipage}{\hsize}},after={\end{minipage}}]
    \vspace*{2pt}
    \item ``The system has integrity'' \cite{jian2000foundations}
    \item ``[...] has integrity, candid, sincere'' \cite{malle2021multidimensional}
    \item ``What \% of the time will this robot [...] tell the truth'' \cite{schaefer2016measuring}
   \end{itemize} \\
   10 & The system uses appropriate methods to carry out operations. & Appropriate &
   \begin{itemize}[before={\begin{minipage}{\hsize}},after={\end{minipage}}]
    \vspace*{2pt}
    \item ``The system uses appropriate methods to reach decisions'' \cite{madsen2000measuring}
    \item ``The system provides very sensitive and effective advice, if needed'' \cite{mcknight2011trust}
    \item ``What \% of the time will this robot… make sensible decisions'' \cite{schaefer2016measuring}
   \end{itemize} \\
   11 & It is difficult for me to identify what the system will do next. & Open (``Openness'') &
   \begin{itemize}[before={\begin{minipage}{\hsize}},after={\end{minipage}}]
    \vspace*{2pt}
    \item ``The system state was always clear to me'' \cite{korber2018theoretical}
    \item ``What \% of the time will this robot [...] clearly communicate'' \cite{schaefer2016measuring}
    \item ``It is difficult to identify what the system will do next'' \cite{korber2018theoretical}
   \end{itemize} \\
   12 & The system provides clear, open information about its current status and actions & Open (``Openness'') &
   \begin{itemize}[before={\begin{minipage}{\hsize}},after={\end{minipage}}]
    \vspace*{2pt}
    \item ``What \% of the time will this robot [...] communicate with people'' \cite{schaefer2016measuring}
    \item ``What \% of the time will this robot [...] openly communicate'' \cite{schaefer2016measuring}
    \item ``What \% of the time will this robot be [...] unresponsive'' \cite{schaefer2016measuring}
   \end{itemize} \\
   13 & The system behaves in ways that are appropriate for the situation. & Appropriate &
   \begin{itemize}[before={\begin{minipage}{\hsize}},after={\end{minipage}}]
    \vspace*{2pt}
    \item ``Steps are typical of other sites'' \cite{gefen2003trust}
    \item ``Interaction is typical'' \cite{gefen2003trust}
    \item ``What \% of the time will this robot [...] follow directions'' \cite{schaefer2016measuring}
   \end{itemize} \\
   14 & The system is designed to act in the best interest of people. & Benevolence &
   \begin{itemize}[before={\begin{minipage}{\hsize}},after={\end{minipage}}]
    \vspace*{2pt}
    \item ``I know it cares about customers'' \cite{gefen2003trust}
    \item ``Vendor has nothing to gain by not caring'' \cite{gefen2003trust}
    \item ``The developers take my well-being seriously'' \cite{korber2018theoretical}
    \item ''Someone you can count on'' \cite{malle2021multidimensional}
   \end{itemize} \\
   15 & I believe the system will make the right decisions when the situation is uncertain. & Faith &
   \begin{itemize}[before={\begin{minipage}{\hsize}},after={\end{minipage}}]
    \vspace*{2pt}
    \item ``When I am uncertain about a decision I believe the system rather than myself'' \cite{madsen2000measuring}
    \item ``To help me perform well, I believe advice from the tank-spotting aid even when I don’t know for certain that it is correct'' \cite{chancey2017trust}
    \item ``If I am not sure about a decision, I have faith that the system will provide the best solution'' \cite{madsen2000measuring}
   \end{itemize} \\
   16 & The system is designed to protect users from harm or negative outcomes. & Protective (``Trustee has a positive orientation towards the trustor [...]'') &
   \begin{itemize}[before={\begin{minipage}{\hsize}},after={\end{minipage}}]
    \vspace*{2pt}
    \item ``The system’s actions will have a harmful or injurious outcome'' \cite{jian2000foundations}
    \item ``What \% of the time will the system [...] protect people'' \cite{schaefer2016measuring}
    \item ``I feel okay using spreadsheet products because they are backed by vendor protections'' \cite{mcknight2011trust}
   \end{itemize} \\
   17 & I believe the system is designed to operate honestly and truthfully toward users. & Honesty (``[...] to what extent the trustee has a motive to lie'') &
   \begin{itemize}[before={\begin{minipage}{\hsize}},after={\end{minipage}}]
    \vspace*{2pt}
    \item ``The system is deceptive'' \cite{jian2000foundations}
    \item ``The system behaves in an underhanded manner'' \cite{jian2000foundations}
    \item ``I know it is honest'' \cite{gefen2003trust}
    \item ``Vendor has nothing to gain by being dishonest'' \cite{gefen2003trust}
    \item ``What \% of the time will the robot [...] tell the truth'' \cite{schaefer2016measuring}
    \item ``sincere, candid'' \cite{malle2021multidimensional}
   \end{itemize} \\
   18 & Even when I'm unsure about the system's decisions, I have faith it is making the right choices. & Faith &
   \begin{itemize}[before={\begin{minipage}{\hsize}},after={\end{minipage}}]
    \vspace*{2pt}
    \item ``Even if I have no reason to expect the system will be able to solve a difficult problem, I still feel certain that it will'' \cite{madsen2000measuring}
    \item ``If I am not sure about a decision, I have faith that the system will provide the best solution'' \cite{madsen2000measuring}
    \item ``If I am not sure about whether to click ``tank present'' or ``tank absent,'' I have faith that the tank-spotting aid will provide the correct solution to help me perform well'' \cite{chancey2017trust}
   \end{itemize} \\
 \bottomrule
\end{longtable}
}

In the taxonomy of \citet{lee2004trust}, Reliable, Functional, Capable, Consistent, Competent, and Predictable relate to ``Performance.''
Understandable, Appropriate, Open, and Integrity relate to ``Process.''
Finally, Benevolence, Faith, Honest, and Protective reate to ``Purpose.''
All these factors can capture aspects of learned trust as subjects in our study passively observe cyber-physical systems.

The table thus grounds our survey instrument as a reasonable interpolation of several prior trust survey instruments in the literature.
In developing this instrument, the authors were encouraged by the consistency---as evidenced by the quotes in the table---of items across trust survey instruments in different domains (from consumer software, to human-robot interaction, to automated systems).
We note that, similar to prior work, several of our items are inverted (i.e., presenting a negative rather a positive statement).
Among other benefits, this helps detect inattentive users who may respond inconsistently with their other answers if they are not paying sufficient attention.
Furthermore, attributes such as Faith and Reliable have two related items, which can also help detect inattentive users if scores on related items are wildly different.
%When computing an overall trust score, the scores for such items are averaged into one sub-score per attribute.

Finally, we note that when conducting a survey for a particular cyber-physical system, we replace all instances of ``the system'' in the survey item with the name of the specific CPS being evaluated---in the case of our user study, ``autonomous vehicles,'' ``power plant,'' or ``robotic assembly line.''

\section{Experiment Videos}
\label{sec:videos}

The videos used in our study (see Section~\ref{sec:user-study}) are publicly available online (note to reviewers: at 30min each, each video is far too large to attach as supplementary material, even with substantial compression):
\begin{enumerate}
    \item \textbf{AV Negative}: \url{https://pxqg-bucket.s3.us-east-2.amazonaws.com/sessions/00374f20-835c-11f0-89ef-910575fc79d5/AV_Negative.mp4}
    \item \textbf{AV Positive}: \url{https://pxqg-bucket.s3.us-east-2.amazonaws.com/sessions/497742f0-83b4-11f0-89ef-910575fc79d5/AV_Positive.mp4}
    \item \textbf{Power Plants}: \url{https://pxqg-bucket.s3.us-east-2.amazonaws.com/sessions/92f20f60-8abb-11f0-9d8d-b174c876e7ab/Power+Plants.mp4}
    \item \textbf{Robotic Assembly Lines}: \url{https://pxqg-bucket.s3.us-east-2.amazonaws.com/sessions/6eb2fbf0-8ab6-11f0-9d8d-b174c876e7ab/Robotic+Assembly+Line.mp4}
\end{enumerate}
%These videos are described in Section~\ref{sec:user-study} in the main body.

% TODO Submit survey instruments, information sheets, etc. as supplementary material in the submission system, not as appendices?  I think?

\end{document}